\documentclass[superscriptaddress,nofootinbib,11pt]{revtex4-1}

\usepackage{graphicx,amsmath,array,verbatim,setspace}
\usepackage{hyperref}       % hyperlinks
\usepackage{url}            % simple URL typesetting
\usepackage{amsfonts}       % blackboard math symbols
\usepackage[italicdiff]{physics}
\usepackage{color}
\usepackage{amssymb,amsthm}
\usepackage{tikz}
\usetikzlibrary{positioning}
\usetikzlibrary{shapes.geometric}
\usetikzlibrary{shapes.misc}
\usepackage{tikz-cd}

\newcommand{\be}{\begin{eqnarray}}
\newcommand{\ee}{\end{eqnarray}}
\newcommand{\nn}{\nonumber}
\newcommand{\nl}{\nonumber \\}
\newcommand{\pd}{\partial}

\def\Z{{\mathbb Z}}
\newcommand{\diff}{\mathrm{d}}
\newcommand{\Diff}{{\mathcal{D}}}
\newcommand{\im}{\mathrm{i}}
\def\coeff#1#2{{\textstyle {\frac {#1}{#2}}}}

\def\half{\coeff 12}

\graphicspath{{./figures/}}

\begin{document}
\title{Exact-WKB, complete resurgent structure, and mixed anomaly\\ in quantum mechanics on $S^1$}
%\title{Exact-WKB analysis for quantum mechanics in a periodic potential}
%\title{Equivalence of  Exact-WKB, resurgent semiclassics, and quantization conditions}
%\title{Unified understanding of exact-WKB and resurgence\\	-\,equivalence of different quantization conditions\,-}

\author{Naohisa Sueishi}
\email{sueishi@eken.phys.nagoya-u.ac.jp}
\affiliation{Department of Physics, Nagoya University, Nagoya 464-8602, Japan}

\author{Syo Kamata}
\email{skamata11phys@gmail.com}
\affiliation{National Centre for Nuclear Research, 02-093 Warsaw, Poland}

\author{Tatsuhiro Misumi}
\email{tatsuhiromisumi@gmail.com}
\affiliation{Department of Mathematical Science, Akita University,  Akita 010-8502, Japan}
\affiliation{Department of Physics, Keio University, Kanagawa 223-8521, Japan}

\author{Mithat \"{U}nsal}
\email{unsal.mithat@gmail.com}
\affiliation{Department of Physics, North Carolina State University, Raleigh, NC 27607, USA}

\begin{abstract}
 We investigate the exact-WKB analysis for quantum mechanics in a periodic potential,  with $N $ minima on $S^{1}$. 
We describe the Stokes graphs of a general potential problem as a network of   Airy-type or degenerate Weber-type building blocks, 
%, and show how the perturbative and non-perturbative cycles are related by using  the cycles of  both the Airy-type ($E_{0}\not= 0$) and the degenerate Weber-type ($E_{0}=0$) Stokes graphs, 
 and provide a dictionary between the two.  The two formulations are equivalent, but  with their own pros and cons. 
% The Weber-type  is more suitable with a merging pair of turning points. 
Exact-WKB produces the quantization condition consistent with the known conjectures and  mixed anomaly.    The quantization condition for the case of $N$-minima on the circle factorizes over the Hilbert sub-spaces labeled by discrete theta angle (or Bloch momenta), and   is consistent with  't Hooft anomaly for even $N$ and global inconsistency for odd $N$. 
%it exhibits two-fold degeneracy over the spectrum at $\theta=\pi$ for even $N$.  This is a result of a mixed  't Hooft anomaly.  
 %between $\Z_N$ translation and charge conjugation symmetry. 
 By using  Delabaere-Dillinger-Pham formula, we prove that 
 the  resurgent structure is closed in these Hilbert  subspaces, built on  discrete  theta vacua,  
% labeled by Bloch momenta, 
 and  by a transformation, this  implies that  fixed topological   sectors (columns of resurgence triangle) are also closed under resurgence. 

\begin{comment}
We investigate the exact-WKB analysis for quantum mechanics in a periodic potential (or quantum mechanics of a particle on $S^{1}$).
We figure out its Stokes graphs and show how the perturbative and non-perturbative cycles are related. The most prominent fact we find in this work is that exact-WKB correctly produces the quantization condition consistent with the known conjecture. 
We also exhibit the dictionary connecting the cycles of the Airy-type ($E_{0}\not= 0$) and the degenerate Weber-type ($E_{0}=0$) Stokes graphs.
By deriving the exact partition function for the periodic-potential quantum mechanics from exact-WKB, we show the detailed resurgent structure. 
We also find that our physical results are consistent with the 't Hooft anomaly for the system.
%We also show that the Weber-type Stokes graph gives exact quantization conditions while the Airy-type one gives approximated condition.
\end{comment}

\end{abstract}

\maketitle

\tableofcontents

\newpage

\section{Introduction}
\begin{figure}[t]
\vspace{-1.5cm}
    \centering
    \includegraphics[scale=0.4]{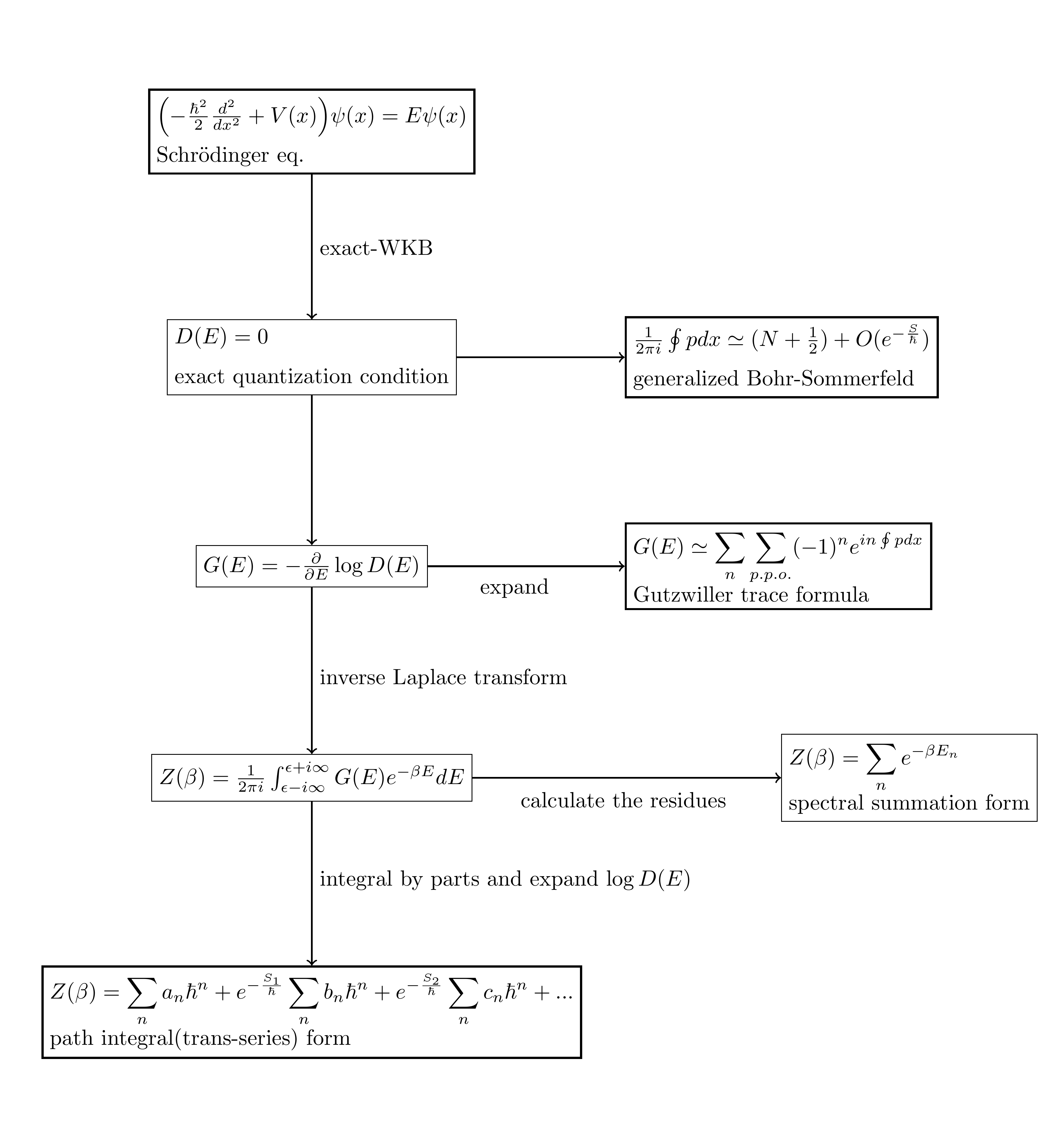}
    \vspace{-10mm}
    \caption{The relation among several quantization methods(${\mathbb Z}_N$-shift symmetry given by $x \rightarrow x+2 p \pi/N$.). We can identify the resurgent structure of each case without approximation from the exact-WKB.}
    \label{fig:flow}
\end{figure}

Recently, the application of resurgence theory and exact-WKB analysis to quantum theory has been attracting a great deal of attention. 
The main statement of the resurgence theory in quantum theory is that the perturbative and non-perturbative contributions have a nontrivial relation 
%through their imaginary ambiguities 
and one can understand many aspects of non-perturbative physics just from the perturbative series. 
The resurgence theory has been intensively investigated in terms of mathematics \cite{Ec1}, quantum mechanics \cite{Brezin:1977ab, Lipatov:1977cd,
Alvarez1,Alvarez2,Alvarez3,
ZinnJustin:2004ib, ZinnJustin:2004cg, Jentschura:2010zza, Jentschura:2011zza, Dunne:2013ada,Basar:2013eka,Dunne:2014bca,Misumi:2015dua,Behtash:2015loa,Gahramanov:2015yxk,Dunne:2016qix,Fujimori:2016ljw,
Basar:2017hpr,Fujimori:2017oab,Sueishi:2019xcj}, matrix models and string theory \cite{Marino:2006hs,Marino:2007te, Pasquetti:2009jg,Garoufalidis:2010ya,Aniceto:2011nu, Marino:2012zq,Aniceto:2013fka,Grassi:2014uua,Couso-Santamaria:2015wga,Hatsuda:2015qzx,Franco:2015rnr,Couso-Santamaria:2016vcc,Couso-Santamaria:2016vwq} and quantum field theory \cite{Dunne:2012ae, Cherman:2013yfa,Cherman:2014ofa,Misumi:2014jua,Misumi:2014rsa,Sulejmanpasic:2016llc,Dunne:2015eaa,Buividovich:2015oju, Gukov:2016njj}. 
The resurgent structure and the related Stokes phenomena are understood by two different methods including semi-classical analysis and the exact-WKB analysis \cite{DDP2,DP1, Takei1,Takei2, Takei3,Kawai1,AKT1, Schafke1,  Iwaki1,Kashani-Poor:2015pca,Kashani-Poor:2016edc,Ashok:2016yxz,Ito:2018eon,Hollands:2019wbr,Ashok:2019gee,Ito:2020ueb,Imaizumi:2020fxf,Coman:2020qgf,Allegretti:2020dyt,Kuwagaki:2020pry,Sueishi:2020rug,Emery:2020qqu,Enomoto:2020xlf,Taya:2020dco,Yan:2020kkb}.
In our previous work \cite{Sueishi:2020rug}, we obtained the unified understanding of the two Stokes phenomena in semi-classical description of path integral  and exact-WKB analyses.  The Stokes phenomenon leading to the ambiguous contribution by the structure of Lefschetz thimble for the quasi-zero mode direction for instanton-antiinstanton critical point at infinity 
corresponds to the change of the "topology" of the Stoke curves in the exact-WKB analysis. 
We also found the relation between Maslov index and the intersection number of Lefschetz thimble. 
The results we obtained in  \cite{Sueishi:2020rug}  is summarized in the flowchart Fig.~\ref{fig:flow}.

In this paper, we study quantum mechanics of a particle  on $S^{1}$  in the presence of  periodic potential. 
%(In this paper, we sometimes call the system ``$S^{1}$ quantum mechanics".) 
We consider $N$-minima on the circle where $N=1, 2, \ldots$.
The application of the exact-WKB analysis to these  systems is of great importance in terms of understanding resurgent  structure in theories with topological $\theta$ angle,  
discrete 't Hooft anomaly \cite{Gaiotto:2017yup}, quantization conditions \cite{ZinnJustin:2004ib,Dunne:2014bca},
the Mathieu equation \cite{Kashani-Poor:2015pca} and TBA equations \cite{Ito:2018eon,Ito:2020ueb,Imaizumi:2020fxf,Emery:2020qqu}. 
Furthermore, these QM systems provide a simpler prototype for circle compactified  $\mathbb {CP}^{N-1}$ on $\mathbb R \times S^1$   \cite{Dunne:2012ae,Sulejmanpasic:2016llc, Misumi:2014jua} 
and deformed Yang-Mills theory on $\mathbb R^3 \times S^1$ \cite{Unsal:2008ch, Unsal:2020yeh}. 
%both of which possess $N$-degenerate perturbative vacua and a $\theta$ angle in their semi-classical calculable regimes. 
We show that the quantization condition   for the system with $N$ minima on  $S^1$ factorizes according to the $N$-Bloch momenta  (or equivalently, $N$ discrete $\theta$ angles), corresponding to decomposition of Hilbert space ${\cal H} = \bigoplus_{p=0}^{N-1} {\cal H}_p $. 
By obtaining the exact partition function of the system based on the exact-WKB analysis, we show that the  resurgent structure  is closed in  
each  ${\cal H}_p$, the eigenspace of the shift operator.   This implies that, by a Fourier transform, the topological sectors $Q \in \Z$ which correspond to columns of resurgence triangle, are also closed under Stokes automorphisms and resurgence. 
We also show that  quantization condition for $N$ even model at $\theta=\pi$ becomes a perfect square, corresponding spectral doubling, and produce 
the mixed  't Hooft anomaly between $\Z_N$ translation symmetry and  $C$ charge conjugation symmetry\cite{Gaiotto:2017yup}.

We elucidate the Stokes graphs of the system. For a classical potential problem,  we describe how Stokes graphs can be expressed as a network of Airy type or degenerate Weber type building blocks.  We show how the perturbative and non-perturbative cycles are related in terms of the resurgent structures. 
The resultant quantization condition is in exact agreement to the conjectured one by Zinn-Justin--Jentschura \cite{ZinnJustin:2004ib} and  Dunne-Unsal \cite{Dunne:2014bca}. 
We also exhibit the dictionary to connect the cycles of the Airy-type ($E_{0}\not=0$) and degenerate Weber-type ($E_{0}=0$) Stokes graphs, where the latter is more suitable for merging pair of turning points. 
%We show that the Weber-type Stokes graph gives exact quantization conditions while the Airy-type one gives approximated conditions.

%%%%%%%%%%%%%%%%%%%%%%

This paper is constructed as follows: In Section \ref{sec:rev}, we review the exact WKB analysis, with emphasis on its relation to the resurgence theory and the known quantization conditions, based on our previous work. In Section \ref{sec:S1}, we study the quantum mechanical systems on $S^{1}$ (periodic-potential systems) by the exact-WKB method with the Airy-type Stokes graph, and obtain the conjectured quantization condition. In Section \ref{sec:H}, we introduce the Hilbert-space perspective and discuss the gauging of ${\mathbb Z}_N$ symmetry, with emphasis on its relation to TQFT. In Section \ref{sec:WB}, we study the $S^{1}$ quantum mechanical systems by the degenerate Weber-type Stokes graph instead of the Airy type, and obtain the quantization condition without any approximation, leading to the partition function with the exact resurgent structure. Section \ref{sec:SD} is devoted to the summary and the discussion.

\subsection{Three related theories} 

There are few  quantum mechanical systems whose local dynamics are identical, but global structure and Hilbert space structures are different. These can be related to each other in a precise way, and our analysis, with some modifications, obviously apply to all three.  For clarity, we briefly describe these three systems and their salient features.

 \begin{itemize}
 \item Particle  on a line  $x \in \mathbb R$  in the  presence of a periodic  potential $V(x+ 2\pi)= V(x)$. This system has a $ \Z$ translation symmetry.  Hilbert space is composed of the  bands and each band has infinitely many states labelled by Bloch momenta $k a  \in [-\pi, \pi]$ (we set lattice spacing $a=1$ in general.) In this construction, there is no theta angle. 
 
\item Gauging $ \Z$ translation symmetry completely,  we end up with particle on a circle,  $x \in S^1 =  \mathbb R/  2 \pi  \mathbb Z $. Now, $x \sim x + 2 \pi$ are physically identified (due to gauging), and there is only one minimum of the potential in the fundamental domain,  $x \in S^1$.     In this system, translation is no longer a global symmetry, it is fully gauged. Only one state from each band of particle on an infinite line $\mathbb R$ is present in the Hilbert space. One can add a theta angle to this system.  Theta angle determines which  Bloch state of the energy band survives in the Hilbert space upon gauging, with identification $ka\equiv \theta$.  We can call the Hilbert space based on this theta vacuum as ${\cal H}_\theta$. 
In the exact WKB analysis,  we generally use this set-up. 

\item Gauging $N  \Z$ subgroup of   $\Z$ translation symmetry,  we end up with particle on a circle,  $x \in S^1 =  \mathbb R/  2 \pi  N  \mathbb Z $. Now, $x \sim x + 2 \pi N$ are physically identified and there are $N$ perturbative minima of the potential in the fundamental domain   $x \in S^1$.   
%We call this system $T_N$ model. 
This system has a genuine global 
$  \Z_N$ translation symmetry. 
 Now,  $N$  states  from each band  are present in the Hilbert space, and these are labelled by $N$ distinct discrete Bloch momenta (also called discrete theta angle in this context).  One can add a continuous theta angle to this system as well.   We use this system in exact WKB analysis  to probe mixed anomalies.  We  call this system $T_N$ model for brevity.
 This set-up can be used to extrapolate between $N=1$ particle on $S^1$ case and particle on an infinite line $x \in \mathbb R$. 
\end{itemize}

These systems possess exactly the same local dynamics. Their perturbation theories, instanton and bion data  are completely equivalent.   But their Hilbert spaces and global symmetries are distinct. As emphasized, the first system has infinitely many states per band, the second system has one state per band, and the third system has $N$ states per band. Yet, one can obtain the whole spectral data of the first system from second and third.  In the exact WKB analysis, we first use the second setup to derive quantization condition for a particular theta 
angle, and build Hilbert space   on  top of a certain theta vacuum  $|\theta \rangle$,   ${\cal H}_\theta$.    We also prove that  quantization condition that produces 
${ \rm Spec}[{\cal H}_\theta]$ is invariant under Stokes automorphism, and DDP formula.  
By a Fourier transform, this shows that traditional resurgence  (relating late terms with early terms) is also closed on the fixed  topological charge sectors, which are the columns of resurgence triangle \cite{Dunne:2014bca}.  Finally, we use the third set-up to demonstrate the  emergence of mixed anomalies at $\theta=\pi$ for even $N$.

\subsection{Review of Exact-WKB and general strategy}
%\subsection{Review of Exact-WKB and Resurgent structure}
\label{sec:rev}

We first briefly review the exact-WKB analysis and its relation to resurgence theory, see  \cite{Sueishi:2020rug} for details.
One of the most important advantages of the exact-WKB analysis is that we obtain the quantization condition from the normalization condition of the wavefunctions in $x\to \pm \infty$ limits of the Stokes graph. The quantization condition is regarded as the zero condition of the Fredholm determinant as
\begin{align}
    D(E)=\det (\widehat{H}-E)=0\,,
\end{align}
where $D(E)$ denotes the Fredholm determinant. 
It enables us to derive exact energy eigenvalues in principle.
Furthermore, once we obtain the Fredholm determinant, we also have the resolvent $G(E)$ and the partition function $Z(\beta)$ straightforwardly as
\begin{align}
    G(E)&=\tr \frac{1}{\widehat{H}-E}=-\pdv{E}\log D(E)\,,
    \\
    Z(\beta)&=\tr e^{-\beta \widehat {H}}=\frac{1}{2\pi i}\int_{\epsilon-i\infty}^{\epsilon+i\infty}G(E)e^{-\beta E}dE\,.
\end{align}
In the previous work \cite{Sueishi:2020rug}, we make use of these facts and show that the Stokes phenomena in the semiclassical path-integral analysis (bion analysis \cite{Unsal:2007vu,Unsal:2007jx,Shifman:2008ja,Poppitz:2009uq,Anber:2011de,Poppitz:2012sw,Misumi:2014raa,Fujimori:2019skd,Misumi:2019upg,Fujimori:2020zka}) are realized as the global alternation of the Stokes graph in the exact-WKB analysis, where the perturbative and nonperturbative contributions correspond to the different cycles crossing the Stokes curves.

In the exact-WKB analysis, the Stokes phenomenon and the related resurgent structure between the perturbative and nonperturbative contributions are determined  by the Stokes curve and its associated monodromy matrix. It is notable that the  Stokes curve is uniquely determined by the lowest order of the WKB expansion, i.e., the  classical potential.

Earlier work  \cite{Sueishi:2020rug} also brought an  understanding of  the relation between the exact-WKB analysis and the other known quantization  methods.
In particular, the trace of resolvent $G(E)$ gives the \textit{Gutzwiller trace formula} \cite{Gutzwiller},
\begin{align}
G(E)=\tr \frac{1}{\widehat{H}-E}=i\sum_{p.p.o.}\sum_{n=1}^\infty T(E)\,e^{in\oint_{p.p.o.} pdx}(-1)^n\qty( \qty| \det \frac{\delta^2S}{\delta x\delta x}|)^{-1/2}\,,
\label{Gutzwiller_form} 
\end{align}
This form is interpreted as the intermediate quantization method between the path integral and the Bohr-Sommerfeld quantizations. This method gives the resolvent of the system by summing up periodic classical solutions.

These facts are summarized in the flowchart shown in Fig.~\ref{fig:flow}.
It is important to note that, since the Fredholm determinant $D(E)$ obtained by the exact-WKB analysis is exact,  what follows from there, 
e.g. the trace  of resolvent $G(E)$ and the partition function $Z(\beta)$  are also exact.

\section{$S^{1}$ quantum mechanical system with Airy-type Stokes graphs}
\label{sec:S1}

\subsection{Quantization condition}

We here discuss the exact-WKB analysis for the  particle on a circle, $x \in S^1$, where $x \sim x + 2\pi$ are physically identified, 
in the presence of the  potential, $V(x)=1-\cos(x)$.  Since the target space is $S^1$, we can turn on topological $\theta$ angle, which correspond to the insertion of the Aharanov-Bohm flux through the circle. 
Our main purpose is to derive the quantization condition from the periodicity condition of the system and WKB-wave function $\psi(x+2\pi)=e^{-i\theta}\psi(x)$. 
%and show its consistency with the known conjecture.
In the sequential subsections, we will obtain the Gutzwiller trace formula of this system, then we will extend our analysis  to the  cases with $V(x)=1-\cos(N x)$, corresponding to $N$-minima in the fundamental domain.

We begin with Schr\"{o}dinger equation
\begin{align}
	\qty(-\frac{\hbar^2}{2}\frac{d^2}{dx^2}+V(x))\psi(x)=E\psi(x)\,. 
\end{align}
Set $Q(x)=2(V(x)-E)$,   rewrite the equation as
\begin{align}     
	\qty(-\frac{d^2}{dx^2}+\hbar^{-2}Q(x))\psi(x)=0\,.
\end{align}
In the WKB analysis, we consider the ansatz given by
\begin{align}
	\psi(x,\hbar) & =e^{\int^x S(x,\hbar)dx} \,,                                     
	\label{WKBpsi}
	\end{align} 
which leads to the 	 to the non-linear  Riccati equation
\begin{align}
	&S(x)^2+\pdv{S}{x}=\hbar^{-2}Q(x) \label{eq:Riccati} 
\end{align}
Next,  we assume that   $S(x,\hbar)$ has  a formal power series expansion in expansion parameter $\hbar$,
\begin{align}
	S(x,\hbar)    & =\hbar^{-1} S_{-1}(x)+S_0(x)+\hbar S_1(x)+\hbar^2 S_2(x)+...\,, 
	\label{WKBS}
\end{align}
where   $S_n(x)$ are functions of $x$.  
This leads to recursive  equation
\begin{align}
	&S_{-1}^2=Q(x)\,,                                                    \quad\quad\quad 2S_{-1}S_n+\sum_{j=0}^{n-1}S_jS_{n-j}+\pdv{S_{n-1}}{x}=0\;\;\;(n\geq 0)\,. 
	\label{Riccati}                     
\end{align}
Since $S_n$ is recursively determined from $S_{-1}=\pm\sqrt{Q}$, $S_n$ has two independent solutions:
\begin{align}
	S^{\pm}(x,\hbar)&=\hbar^{-1} S^{\pm}_{-1}(x)+S^{\pm}_0(x)+\hbar S^{\pm}_1(x)+\hbar^2 S^{\pm}_2(x)+...\nl
	&=\pm\hbar S_{-1}^+ +S_0^+ \pm \hbar S_1^+ +\hbar^2 S_2^+ +..\nl
	& =\pm S_{{\rm odd}}+S_{{\rm even}}\,. \label{eq:SodSev} 
\end{align}
Then the WKB wave functions can be expressed as
\begin{align}
	\psi^\pm_a(x)=e^{\int^x S^\pm dx} =\frac{1}{\sqrt{S_{{\rm odd}}}}e^{\pm\int^x_{a} S_{{\rm odd}} dx}\,, 
\end{align}
with $a$ being an integral constant.
For later calculations, we choose it as a turning point, which is a solution of $Q(x)=0$.

Since we have derived the WKB wave function recursively, it is regarded as a formal series in  $\hbar$ 
\begin{align}
	\psi^\pm_a(x) & =e^{\pm\frac{1}{\hbar}\int_{a}^x \sqrt{Q(x)}dx}\sum_{n=0}^{\infty}\psi_{a,n}^\pm(x)\hbar^{n+\frac{1}{2}}\,, 
	\\
	S_{{\rm odd}}       & =\sum_{n=0}^\infty S_{2n-1}\hbar^{2n-1}\,.                                                                  
\end{align}
Here, both of these series turn out to be asymptotic expansions with respect to $\hbar$. The exact-WKB analysis considers the Borel summation of each series and their Stokes phenomena \cite{Sueishi:2020rug}. 

\begin{figure}[t]
    \centering 
    \includegraphics[width=6cm]{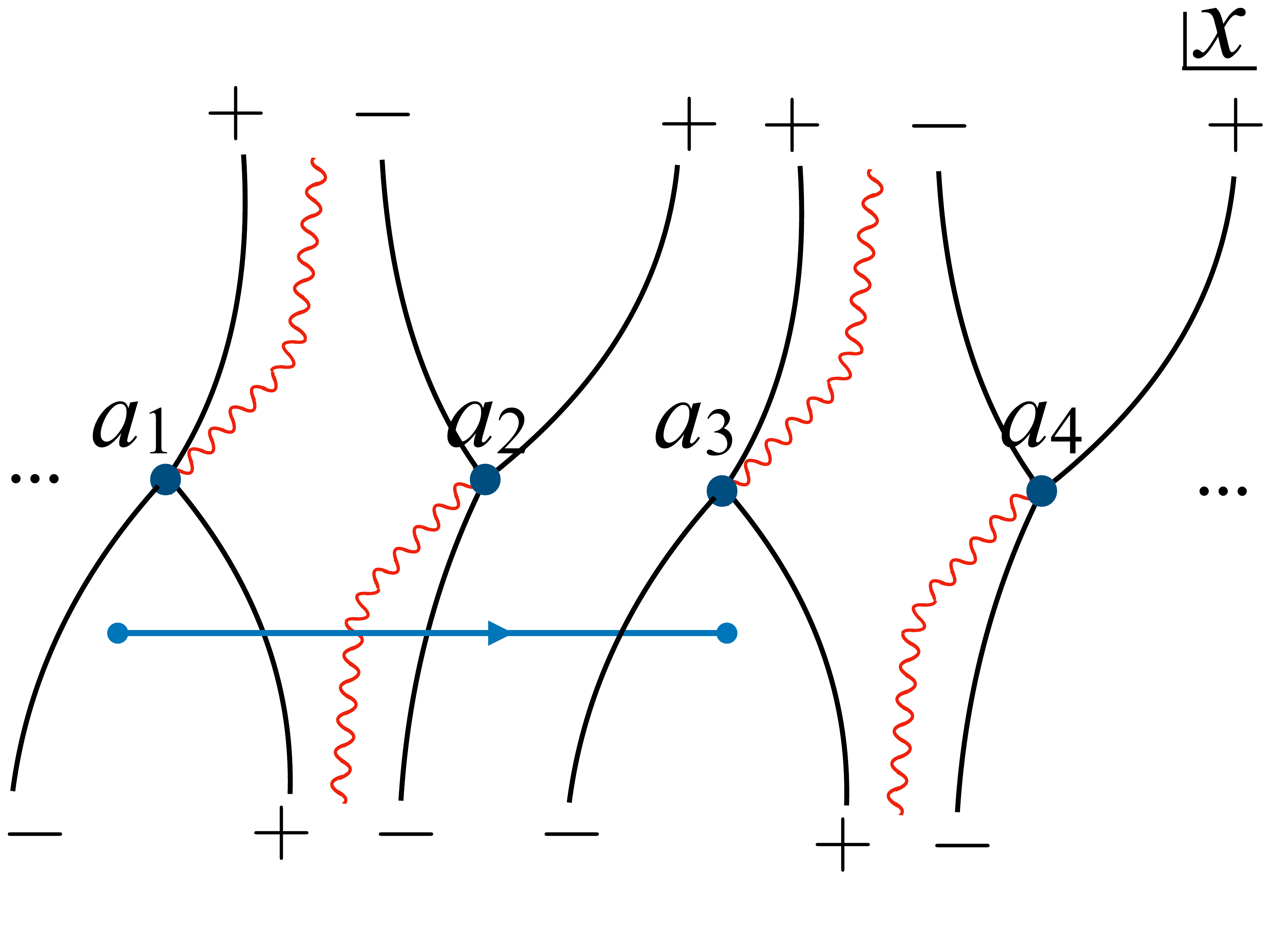}
    \quad\quad\quad\quad\quad\quad
    \includegraphics[width=6cm]{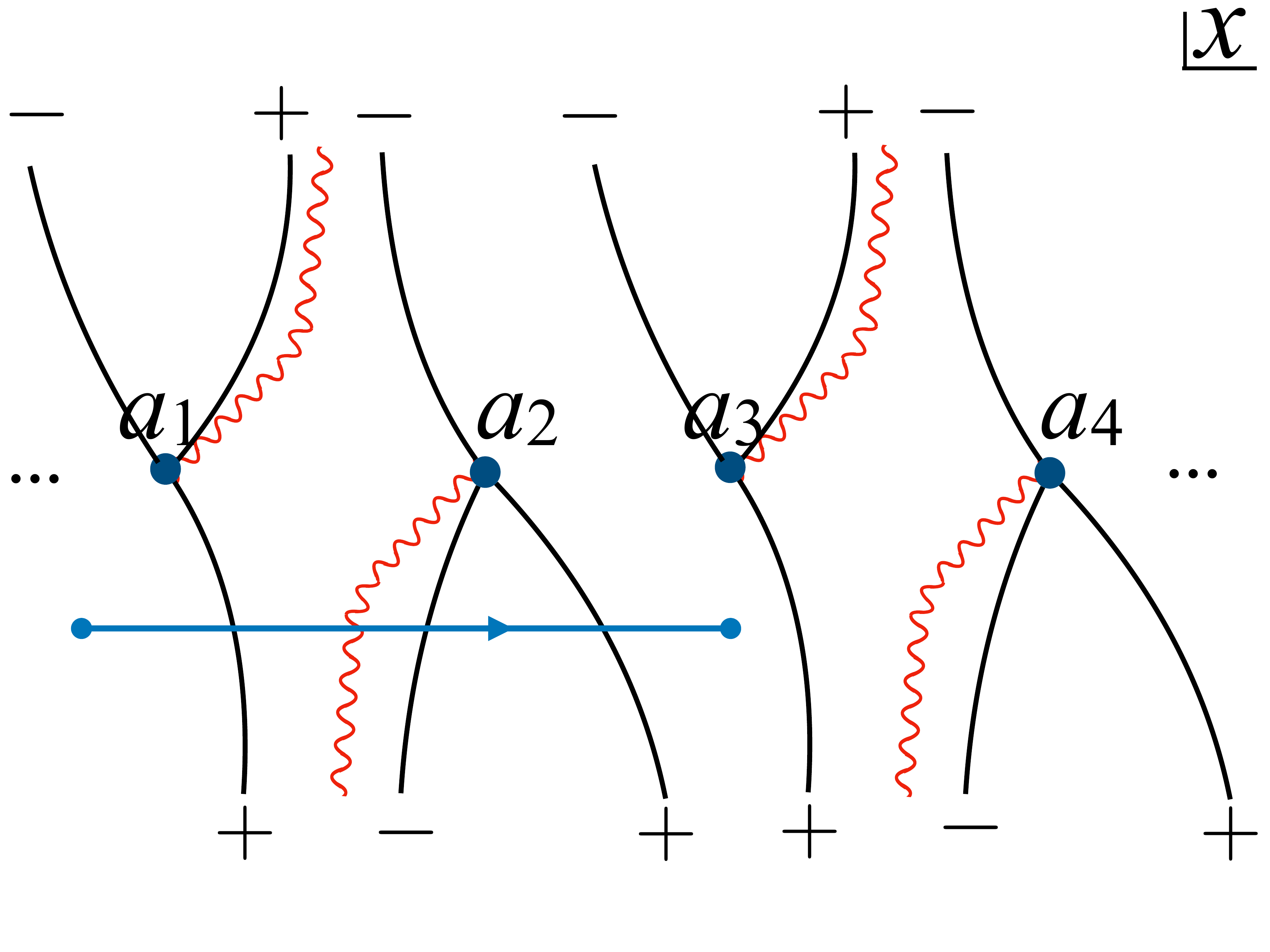}
    \caption{The Stokes curve for the two periods of the potential $1-\cos(x)$ for ${\rm Im} (\hbar) >0$ and  ${\rm Im} (\hbar) <0$, respectively.   We also depict the branch cut, the turning points and the path corresponding to the single period.}
    \label{fig:S1}
\end{figure}

\begin{figure}[t]
    \centering 
    \includegraphics[width=6cm]{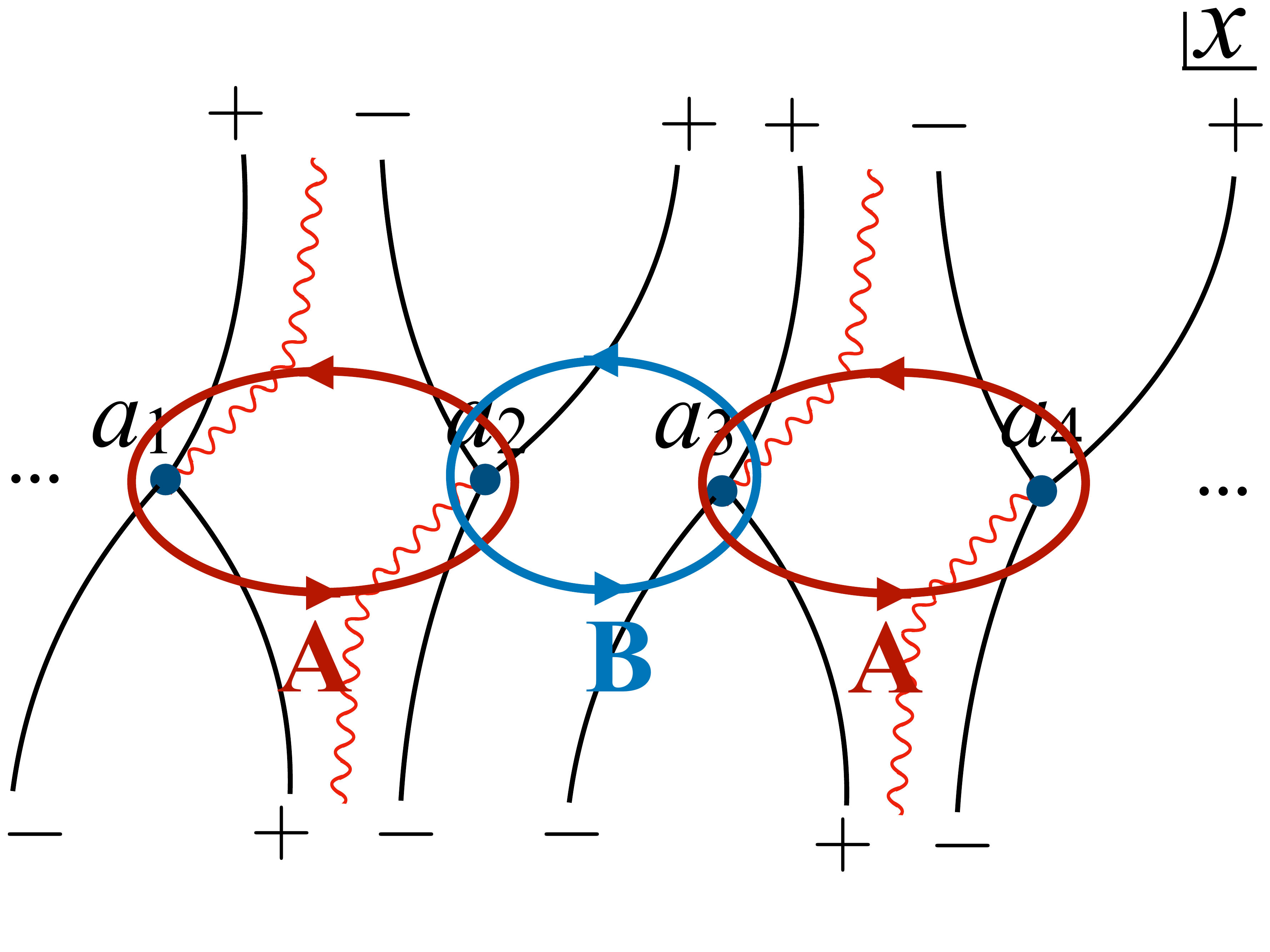}
    \quad\quad\quad\quad\quad\quad
    \includegraphics[width=6cm]{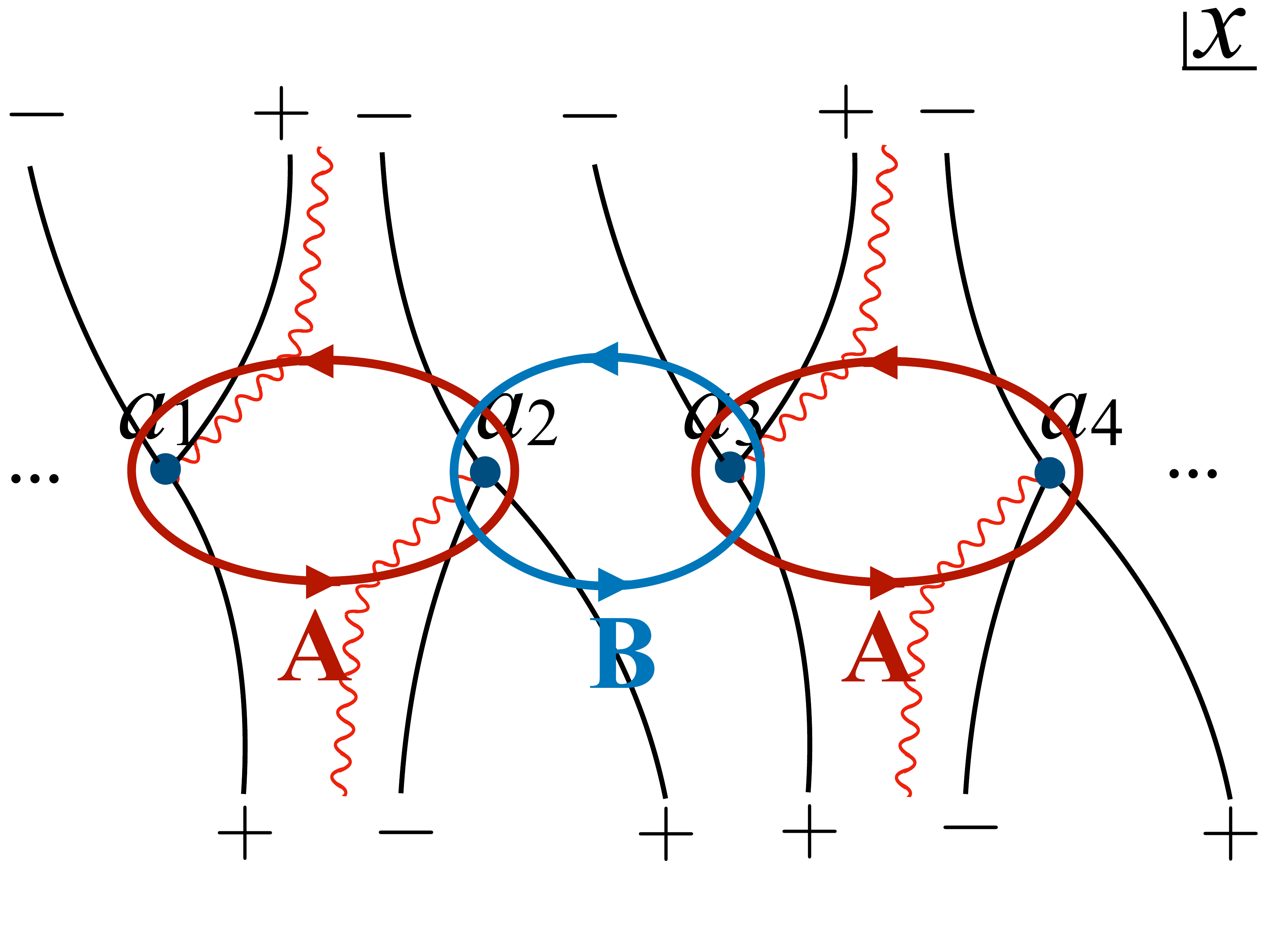}
    \caption{The Stokes curve for the two periods of the potential $1-\cos(x)$, with the cycles $A$ and $B$ being depicted.}
    \label{fig:S1-cycle}
\end{figure}

From now  on, we focus on  the periodic  potential $V(x)=1-\cos(x)$, and  $x\in S^1$. 
%or equivalently the periodic potential. As a typical case, we adopt the periodic potential $V(x)=1-\cos(x)$.
We now determine  the Stokes curve, which dictates  where the Stokes phenomenon of WKB wave function occurs. 
Let $a$ be a turning point (a solution of $Q(x)=0$). The Stokes curve associated with $a$ is defined as
\begin{align}
	\Im \frac{1}{\hbar}\int_a^x \sqrt{Q(x)} dx=0 \,.
\end{align}
Each  segment of the  Stokes curve has an index, $\pm$.  This index indicates which one of the  $\psi^+$ and $\psi^-$  pair increases exponentially when  moving from the point $a$ to infinity along the Stokes curve. 
When the index of the corresponding Stokes curve is $+$,  $\psi^+$ increases exponentially with 
\begin{align}
	\Re \frac{1}{\hbar}\int_a^x \sqrt{Q(x)} dx >0 \,.
\end{align}
When  the index is $-$, then  $\psi^-$ increases exponentially with
\begin{align}
	\Re \frac{1}{\hbar}\int_a^x \sqrt{Q(x)} dx <0 \,.
\end{align}
In the present case, the Stokes curve is depicted in Figs.~\ref{fig:S1} and \ref{fig:S1-cycle}.
In Fig.~\ref{fig:S1} we depict Stokes curves for the two periods of the potential,  exhibiting  four turning points $a_{1}, a_{2}, a_{3}, a_{4}$ and the path corresponding to a single period by a blue line.
In Fig.~\ref{fig:S1-cycle} we also exhibit the perturbative cycle $A=e^{\oint_A S_{{\rm odd}}}=e^{2\int_{a_1}^{a_2}S_{{\rm odd}}} = e^{2\int_{a_3}^{a_4}S_{{\rm odd}}}$\,,
and 
the non-perturbative cycle
$B=e^{\oint_A S_{{\rm odd}}}=e^{2\int_{a_2}^{a_3}S_{{\rm odd}}}$\,,
in the same figure. 
The nonperturbative cycle $B$ corresponds to the single bion contribution $\propto e^{-S_{\rm bion}/\hbar}$.
We note that, although this Stokes curve is specific to the potential $V(x)=1-\cos(x)$, any periodic potential has  one with the same topological property.

Based on the Stokes graph, we derive the quantization condition below for the present potential.
We here consider the single-period path depicted in Fig.~\ref{fig:S1-cycle}.
What we have to do is just to find out the monodromy matrices appearing when the path is  crossing the Stokes curves. 
As we cross the Stokes graph through the full period, 
the monodromy matrices for $\Im \hbar>0$ are given as
\begin{align}
	\mqty(\psi^+_{a_1}(x) \\ \psi^-_{a_1}(x))&=M_+TN_{a_1a_2}M_-N_{a_2a_3}M_-\mqty(\psi^+_{a_3=a_1}(x+2\pi) \\ \psi^-_{a_3=a_1}(x+2\pi))\\
	& \equiv\mathcal{M^+}\mqty(\psi^+_{a_1}(x+2\pi) \\ \psi^-_{a_1}(x+2\pi))\,,
\end{align}
and these for $\Im\hbar<0$ are given as
\begin{align}
	\mqty(\psi^+_{a_1}(x) \\ \psi^-_{a_1}(x))&=M_+TN_{a_1a_2}M_-M_+N_{a_2a_3}\mqty(\psi^+_{a_3=a_1}(x+2\pi) \\ \psi^-_{a_3=a_1}(x+2\pi))\\
	& \equiv\mathcal{M^-}\mqty(\psi^+_{a_1}(x+2\pi) \\ \psi^-_{a_1}(x+2\pi))\,,
\end{align}
Here, $M_{\pm }$ acts while passing $\pm$ labelled  Stokes line in the  counter-clockwise direction, $T$ acts on the crossing of the branch cut in the counter-clock-wise direction, $N_{a_1 a_2}$ is the Voros multiplier accounting the change of turning points entering the WKB wave-function. These are explicitly given by:
\begin{align}
    M_{+}&:=\mqty(1 && i\\ 0 && 1),\quad\quad\quad M_{-}:=\mqty(1 && 0\\ i && 1)\nl
    T&:=\mqty(0 && -i\\ -i && 0),\quad N_{a_1 a_2}:=\mqty(e^{+\int_{a_1}^{a_2}S_{{\rm odd}}} & 0 \\ 0 & e^{-\int_{a_1}^{a_2}S_{{\rm odd}}})\,.
\end{align}

We now impose a boundary  condition on the above wave-function specific to the periodic potential.
Because of the $2\pi$ periodicity of $x$, the wave-function must  satisfy $\psi(x+2\pi)=e^{-i\theta}\psi(x)$. 
We then have the condition
\begin{align}
    \mathcal{M}^\pm\mqty(\psi^+_{a_1}(x) \\ \psi^-_{a_1}(x))=e^{i\theta}\mqty(\psi^+_{a_1}(x) \\ \psi^-_{a_1}(x))\,.
\end{align}
This is nothing but the eigenvalue equation of $\mathcal{M}^\pm$. Therefore, we obtain
\begin{align}
    \det (\mathcal{M}^\pm-e^{i\theta}I)=0,
\end{align}
where $I$ is a $2\times 2$ unit matrix.
\begin{comment}
\begin{align}
    D^+&=\frac{1}{\sqrt{B}}\qty(\frac{1}{\sqrt{A}}+\sqrt{A})+\sqrt{B}\frac{1}{\sqrt{A}}-2\cos\theta=0\\
    D^-&=\frac{1}{\sqrt{B}}\qty(\frac{1}{\sqrt{A}}+\sqrt{A})+\sqrt{B}\sqrt{A}-2\cos\theta=0\\
\end{align}
\end{comment}
This result means that the Fredholm determinant $D(E)$ in the quantization condition $D(E)=0$ is $D^\pm=\frac{1}{e^{i\theta}} \det (\mathcal{M}^\pm-e^{i\theta}I)$, where $\pm$ indicates the sign of imaginary term of $\hbar$.
We now write down the quantization condition for the present periodic potential as\footnote{The overall constant doesn't affect the quantization condition and chosen for simplicity.}
\begin{align}
    D^\pm(E)=\frac{1}{\sqrt{A^{\mp1}B}}\qty(1+A^{\mp 1}+A^{\mp 1}B-2\sqrt{A^{\mp 1}}\sqrt{B}\cos\theta)=0\,.
    \label{eq:D_S1}
\end{align}
This  result %is one of our main results, and 
agrees with  \cite{Kashani-Poor:2015pca} where quantization condition for Mathieu equation was obtained by use of the exact-WKB method as well.   For our purpose,  \eqref{eq:D_S1} is a building block, as it will become manifest in our treatment of    
a potential with $N$-minima instead of one in the fundamental domain, as discussed   in Sec.~\ref{subsec:Nx}.  This generalization will allow us to make a precise link between exact WKB method and mixed 't Hooft anomalies. Furthermore, we will use  \eqref{eq:D_S1}  to prove that fixed topological charge sectors of the theory (corresponding to the columns of resurgence triangle)  are closed under  Stokes automorphism.

%Here, we make comments on the relation between our results and those in the reference \cite{Kashani-Poor:2015pca}. The quantization condition for Mathieu equation was obtained by use of the exact-WKB method in \cite{Kashani-Poor:2015pca} too. In our work, however, we derive the quantization condition for more generic periodic potential as $1-\cos (Nx)$ in Sec.~\ref{subsec:Nx} and obtain the expression of the  partition function with revealing the complete resurgent structure in Sec.~\ref{sec:WB}. Thus, our results are obviously beyond what has been already done.

The resurgent structure of $D^{\pm}(E)$ is determined by Delabaere-Dillinger-Pham (DDP) formula \cite{DDP2}\cite{Iwaki1}. 
\begin{align}
    \mathcal{S}_+[\sqrt{A}]&=\mathcal{S}_-[\sqrt{A}](1+\mathcal{S}[B])
    \label{eq:DDP_fom}
\end{align}
where $\mathcal{S}_\pm$ is directional/lateral Borel summation\footnote{The Borel summation is a homomorphism, so that the following algebraic properties hold: 
$\mathcal{S}[A+B]=\mathcal{S}[A]+\mathcal{S}[B], \mathcal{S}[AB]=\mathcal{S}[A]\mathcal{S}[B]$} for sign($\Im \hbar)=\pm1$. ($B$ cycle does not have Borel singularity so $\mathcal{S}_+[B]=\mathcal{S}_-[B]=\mathcal{S}[B]$.) The DDP formula states that the left/right Borel resummation of the perturbative $A$-cycle is dictated by the   Borel resummation of the non-perturbative $B$ cycle. 
Using DDP formula, we can show that left/right Borel resummation of the exact quantization condition are equal: 
\begin{align}
    \mathcal{S}_+[D^+]& = \mathcal{S_-}[D^-].
    \label{invar}
\end{align}
Therefore, the Fredholm determinant is  invariant under the change of   directions (left or right) in the  Borel summation or equivalently, under Stokes automorphism. 

We can show a physical meaning of this condition.
To show the non-perturbative contribution to the ground state energy, 
we consider the asymptotic form of $A$, which does not include non-perturbative contribution 
before being Borel-resummed.
It is written as
\begin{align}
  A \rightarrow e^{-2\pi i \frac{E}{\hbar \omega_A(E,\hbar)}}\,,
  \label{eq:Aomega1}
\end{align}
where $\omega_A(E,\hbar)$ is an asymptotic expansion with respect to $\hbar$. 
In the low-energy limit, it is regarded as a harmonic frequency of the classical vacuum as
\begin{align}
  \omega_A(E,\hbar)^2&=\sum_{n=0}^\infty c_n(E)\hbar^n\\
  \lim_{E\rightarrow 0}c_0(E)&=V''(x_{\rm vac})\,,
  \label{eq:Aomega2}
\end{align}
where $x_{\rm vac}$ is a minimum of the potential. 
This expression corresponds to taking the Borel-resummed $A$ back to its asymptotic expansion form. 
We now express the energy eigenvalues as $E=\hbar\omega_A(\frac{1}{2}+\delta)$.
Then, the non-perturbative energy deviation $\delta$ from the harmonic oscillator is 
\begin{align}
    \sin(\pi\delta)=\pm i\frac{1}{2}Be^{\pm\pi i\delta}-\sqrt{B}\cos\theta\,,
    \label{eq:sindel}
\end{align}
and it is approximated as
\begin{align}
    \delta\sim -\frac{1}{\pi}\sqrt{B}\cos\theta\pm i\frac{1}{2\pi}B\,.
    \label{eq:del}
\end{align}
We note that $\sqrt{B}$ corresponds to the instanton contribution $\propto e^{-(S_{\rm bion} /2)/\hbar} = e^{-S_{I}/\hbar}$, with $S_{I}$ being the instanton action.
One finds that this expression of the ground state energy indicates the instanton contribution shifts the energy by an amount dictated by $\theta$ angle,  
while the bion contribution leads to the imaginary ambiguity associated with the Stokes phenomena. The ambiguity cancels against the  Borel resummation 
of perturbation theory. The meaning of \eqref{invar} is that this type of resurgent cancellation takes place to all non-perturbative orders. 

Despite the elegance of the Airy-type analysis of exact-WKB,  this formalism is not always  most suitable, especially  when the turning points merge. This  limit requires  some extra work to get the spectral information correctly, and this task does not seem to be very insightful.  Instead, we  discuss the method of degenerate Weber-type exact-WKB. The formalism  we already built-in for Airy will be quite useful there, and we will also provide a dictionary between these  two types of exact-WKB.   Weber-type exact-WKB produce spectral data correctly as discussed in   section \ref{sec:WB}.

\begin{comment}
For $\theta=\frac{\pi}{2}$, the instanton contribution to the ground state energy disappears.
Our result of the ground state energy is consistent with the conjecture given by Zinn-Justin-Jentschura \cite{ZinnJustin:2004ib} and Dunne-Unsal \cite{Dunne:2013ada}.
Eqs.~(\ref{eq:D_S1})(\ref{eq:sindel})(\ref{eq:del}) shows both the nontrivial resurgent structure and the correctness of the conjectured quantization condition for the $S^{1}$ quantum mechanical system just from the Schr\"{o}dinger equation.
To obtain the precise form of eigenvalues, it is convenient to use the degenerate Weber-type exact-WKB instead of the Airy-type analysis used here. We discuss the method of degenerate Weber-type exact-WKB and the eigenvalue in detail in section \ref{sec:WB}.
\end{comment}

\subsection{Gutzwiller trace formula}
The Gutzwiller trace formula is a semiclassical construction that express the quantum mechanical density of states (the resolvent,  $G(E)$),
in terms of periodic orbits, which is called  prime periodic orbit (p.p.o.).
It is generally difficult to determine what the p.p.o.  are and how they are added. But   this data   can be easily extracted by using the quantization condition obtained by exact-WKB, as shown in \cite{Sueishi:2020rug}. In this subsection, we show the structure of Gutzwiller trace formula of $S^1$ system and how Stokes phenomenon corresponding to the  imaginary term of $\hbar$ appears in this formalism. 

\begin{align}
    D^\pm&\propto 1+A^{\mp 1}+A^{\mp}B-2(\sqrt{A})^{\mp 1}\sqrt{B}\cos\theta\nl
    &=(1+A^{\mp 1})\qty(1+\frac{B}{1+A^{\pm 1}}-\frac{\sqrt{B}}{\sqrt{A}+\frac{1}{\sqrt{A}}}(e^{i\theta}+e^{-i\theta}))
\end{align}
Using $G^{\pm}(E)=-\pdv{E}\log D^\pm$, 
\begin{align}
    G^{\pm}_{\rm pt}(E)&=-\qty(\pdv{E}A^{\mp 1})\sum_{n=0}^\infty (-1)^n A^{\mp n}\nl
    G^{\pm}_{\rm np}(E)&=-\qty(\pdv{E}K)\sum_{m=0}^\infty(-1)^mK^m\nl
    K&=B\sum_{n=0}^\infty (-1)^nA^{\pm n}-\sqrt{B}\sum_{n=0}^\infty(-1)^nA^{\pm \qty(n+\frac{1}{2})}(e^{i\theta}+e^{-i\theta})
    \label{eq:Gutzwiller}
\end{align}

\begin{figure}[t]
    \centering 
    \includegraphics[width=8cm]{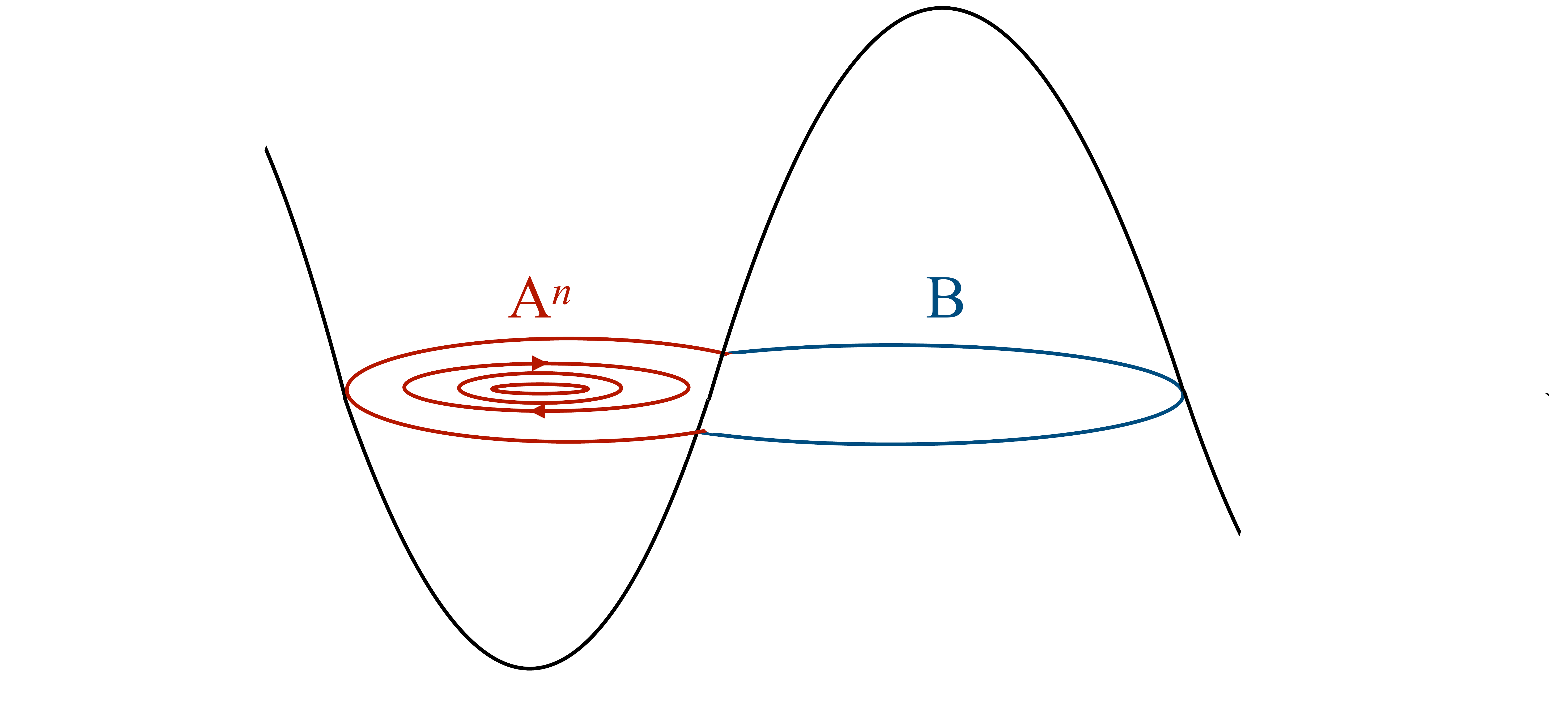}
    \quad
    \includegraphics[width=8cm]{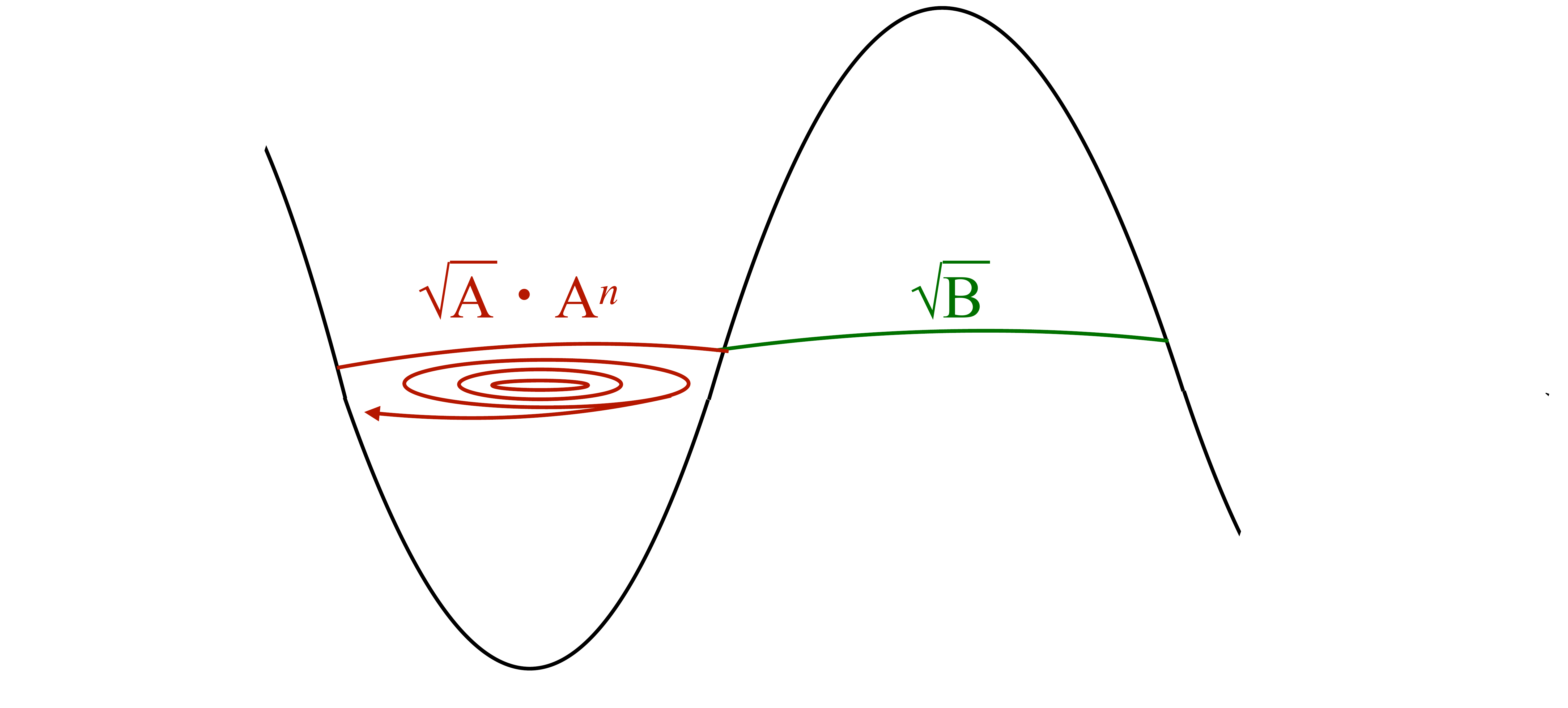}
    \caption{To be periodic, $B A^n$ and $\sqrt{B}A^{n+\frac{1}{2}}$ is the unit of nonperturbative contribution. The Stokes phenomenon corresponding to the bion ambiguity is regarded as $A\rightarrow A^{-1}$}
\end{figure}
Now $A=e^{\oint_A S_{odd}dx}\simeq e^{\frac{i}{\hbar}\oint_A |p|dx}$ and $B\simeq e^{-\frac{1}{\hbar}\oint_B |p|dx}$. From (\ref{eq:Gutzwiller}), we can identify the Gutzwiller trace formula and p.p.o., which includes the non-perturbative contribution in a periodic system. The derivative term $\pdv{E}A=\frac{1}{\hbar}(\oint_A \frac{1}{\sqrt{2(V(x)-E)}}dx+O(\hbar))A=\frac{i}{\hbar}T_A A$ gives the period\footnote{This period includes quantum fluctuation $O(\hbar)$. The original Gutzwiller trace formula is derived with semi-classical approximation, so such the fluctuation term (including the higher order in  $A$, $B$) gives the correction for the Gutzwiller trace formula.}. The $(-1)^n$ associated with each periodic orbit is the Maslov index\footnote{More precisely, the Maslov index is $\alpha$, where $(-1)^n=e^{i\alpha \pi}$}. 
The physical meaning of the form of $K$, which is the unit of the non-perturbative contribution, can also be understood by considering its orbit. There are two kinds of fundamental nonperturbative periodic orbits, $BA^n$ and $\sqrt{B}A^{n+\frac{1}{2}}$ in the periodic potential as shown in Fig.\ref{fig:S1-cycle}. Actually, the infinite number of $A$ cycle attached to $B$ or $\sqrt{B}$ in the expression of $K$  is regarded as quasi-moduli integral  in terms of the path integral method \cite{Sueishi:2020rug} and we show it explicitly in \ref{Sec:Partition}.

%%%%%%%%%%%%%%%%%%%%%%%%%%%%%%%%
\subsection{For $V(x)=1-\cos(Nx)$}
\label{subsec:Nx}
For more generic cases $V(x)=1-\cos(N x)$ ($N \in {\mathbb N}.$), we can also obtain the Fredholm determinant and the quantization condition.
The monodromy matrix unit $\mathcal{M}^{\pm}$ is diagonalized as
\begin{align}
    U^{-1}\mathcal{M^\pm}U=\mqty(\xi-\sqrt{\xi^2-1} && 0 \\ 0 && \xi+\sqrt{\xi^2-1})=\mqty(\alpha && 0 \\ 0 && \beta)\, ,
\end{align}
with $\xi=\frac{1}{2\sqrt{A^{\pm 1}B}}(1+A^{\pm 1}+B)$.
Here $\alpha, \beta$ are the roots of $x^2-2\xi x +1=0$, where we have $\alpha+\beta=2\xi, \alpha\beta=1$.
We then have the quantization condition as
\begin{align}
    \det (({\mathcal{M}}^\pm)^{N} - e^{i\theta}I)&=\det ((U^{-1}\mathcal{M}^\pm U)^{N}- e^{i\theta}I)\nl
    &=e^{i\theta}(2\cos\theta-\alpha^{N}-\beta^{N}) = 0\,,
\end{align}
which is rewritten as
\begin{align}
    D^\pm(E)&=\alpha^{N}+\beta^{N}-2\cos\theta\nl
    &=\Bqty{2\sum_{\ell=0}^{\lfloor \frac{N}{2} \rfloor}\mqty(N\\N-2\ell)\qty(\frac{1+A^\pm+B}{2\sqrt{A^\pm B}})^{N-2\ell}\qty(\frac{(1+A^\pm+B)^2}{4A^\pm B}-1)^\ell}^2-2\cos\theta\nl
    &=\frac{1}{({A}^{\mp 1} {B})^{N/2}} \prod_{p=0}^{N-1} \left[ 1 +{A}^{\mp 1} \left(1 +  {B} \right) - 2 \sqrt{{A}^{\mp 1}{B}} \cos \left( \frac{ \theta + 2 \pi p}{N}\right) \right] \nl 
 & \equiv  \prod_{p=0}^{N-1}  D_{p}^\pm(E) 
    \label{eq:Airy_N}
\end{align}
where $\lfloor x \rfloor$ is the floor function and $\mqty(N\\N-2\ell)$ is the binomial coefficient.
This is one of the most important results in this paper. First, it shows that the exact quantization condition factorizes to N-building blocks. These building blocks are labelled by a Bloch momentum (discrete theta angle). This factorization is due to the fact that Hilbert space decompose to eigenstates of $\Z_N$ translation operator, and we explain the details of this in the next section.  

When $N=2K \;\;(K\in {\mathbb N})$ and $\theta=\pi$, the exact quantization  condition \eqref{eq:Airy_N}  becomes a  perfect square:
\begin{align}
    D(E)&=(\alpha^K+\beta^{K})^2\nl
    &=\frac{1}{({A}^{\mp 1} {B})^{K}} \prod_{p=0}^{K-1} \left[ 1 +{A}^{\mp 1} \left(1 +  {B} \right) - 2 \sqrt{{A}^{\mp 1}{B}} \cos \left( \frac{ \pi (2p+1)}{2K}\right) \right]^2,
\end{align}
This  indicates that all the energy eigenvalues are doubly degenerate (Kramers doubling).
This degeneracy is regarded as a result of 't Hooft anomaly between 
$\Z_N$ discrete translation symmetry and $C=\Z_2$ charge conjugation symmetry.  
The existence of 't Hooft anomaly means that a trivial gap with a unique ground state is prohibited. 
Thus, the ground state should be degenerate if a mixed  't Hooft anomaly exists  of the quantum mechanical models. 
In the next section, we will discuss the gauging of ${\mathbb Z}_N$ symmetry, and the result of the mixed 't Hooft anomaly in detail.

When $N=2K +1 \;\;(K\in {\mathbb N})$ and $\theta=0, \pi$, the exact quantization  condition \eqref{eq:Airy_N}  becomes 
\begin{align}
    D(E, \theta=0)&=    D_{p=0}(E,\theta=0)  \prod_{p=1}^{K}   \Big [D_p(E,\theta=0) \Big ]^2   \cr 
    D(E, \theta=\pi)&=    D_{p=K}(E, \theta=\pi)  \prod_{p=0}^{K-1}    \Big [D_p(E,\theta=\pi)  \Big ]^2 \; .
\end{align}
 There are $K$ pairs, and a singlet sector.  The sector that is not paired up at $\theta=0$ and $\theta=\pi$ are distinct, and they are not continuously connected.   This is the global inconsistency condition \cite{Kikuchi:2017pcp}. 
It is slightly milder condition than mixed anomaly, but essentially plays similar role.  Exact quantization naturally captures global 
inconsistency condition  as well.

\begin{comment}
When $N=2K \;\;(K\in {\mathbb N})$ with $\theta=\pi$, the condition is written as the square perfect
\begin{align}
    D(E)&=(\alpha^K+\beta^{K})^2\nl
    &=\frac{1}{({A}^{\mp 1} {B})^{K}} \prod_{p=0}^{K-1} \left[ 1 +{A}^{\mp 1} \left(1 +  {B} \right) - 2 \sqrt{{A}^{\mp 1}{B}} \cos \left( \frac{ \pi (2p+1)}{2K}\right) \right]^2,
\end{align}
This square perfect indicates that all the eigenvalues are doubly degenerate.
This is again the result of the 't Hooft anomaly.
In the next section, we will discuss the gauging of ${\mathbb Z}_N$ symmetry and the result of the mixed 't Hooft anomaly in detail.
\end{comment}

%%%%%%%%%%%%%%%%%%%%%%%%%%%%%%
\section{Hilbert space perspective and ${\mathbb Z}_N$ gauging}
\label{sec:H}

In this section we introduce the Hilbert-space perspective and interpret  our result of the quantization condition in Eq.~(\ref{eq:Airy_N}).
We also discuss the gauging of ${\mathbb Z}_{N}$ symmetry and show that the mixed 't Hooft anomaly is encoded in our result in Eq.~(\ref{eq:Airy_N}).

\subsection{Factorization of exact quantization condition from Hilbert space perspective}
Consider quantum mechanics of a particle on a circle 
with the potential 
\begin{align} 
V(x) &= 1 - \cos(Nx),  \qquad x \sim x+ 2\pi 
\label{pot}
\end{align} 
For brevity, we call it  $T_N$ model \cite{Unsal:2012zj}. This theory has a discrete $\Z_N$ translation symmetry,  
 \begin{align} 
\Z_N:  x  \mapsto  x+  \frac{2\pi}{N} 
\label{sym}
\end{align}
whose generator we denote with  
$\mathsf {U}$.   Since $[H, \mathsf {U}]=0$, eigenstates of Hamiltonian are also eigenstates of  discrete translation operator.  Denote eigenstates as 
$|n, p \rangle $ where $n$ is band label and $p$ is the label of  the  Bloch momentum associated with  $\Z_N$ symmetry, 
obeying  
\begin{align}
H |n, p \rangle = E_{n, p} |n, p \rangle,  \qquad   \mathsf {U}  |n, p \rangle =  e^{ i 2\pi p /N}  |n, p \rangle    
\label{band}
\end{align}
The $\mathsf {U}$ operator obeys   $\Z_N$  group multiplication law: 
 \begin{align}
 \mathsf {U}^{\ell_1}   \mathsf {U}^{\ell_2}=  \mathsf {U}^{\ell_1  +  \ell_2 \; {\rm mod } \; N}
 \end{align}   
Given the  $\mathsf {U}$ operator, we can built projection operator  to Bloch momentum $p$ states that is useful to decompose  the Hilbert space into 
Bloch sectors: 
\begin{align}
\Pi_p =  \frac{1}{N} \sum_{\ell =0}^{N-1} \omega^{\ell p} \mathsf {U}^\ell 
\label{proj}
\end{align}   
The projection operators satisfy  the standard relations:
\begin{align}
 \sum_{p=0}^{N-1} \Pi_p =  {\bf 1},   \qquad \Pi_p^2 = \Pi_p, \qquad \Pi_{p_1}  \Pi_{p_2} = 0  \;   {\rm if }  \; p_1 \neq p_2  \; {\rm mod } \; N
 \end{align}  
 and can be used to   decompose the Hilbert space of the theory into sectors
\begin{align}
{\cal H}=\bigoplus_{p=0}^{N-1}   {\cal H}_p 
\label{decom}
\end{align}
according to Bloch momenta.   This decomposition is one reason for the factorization of  the exact quantization condition \eqref{eq:Airy_N}. Relatedly,    $ {\cal H}_p $ subspaces  in this decomposition  will emerge naturally as we gauge $\Z_N$ symmetry, as the full Hilbert space of  $(T_N/\Z_N)_p$ models.

\begin{figure}[t]
\begin{center}
\includegraphics[width = 0.8 \textwidth]{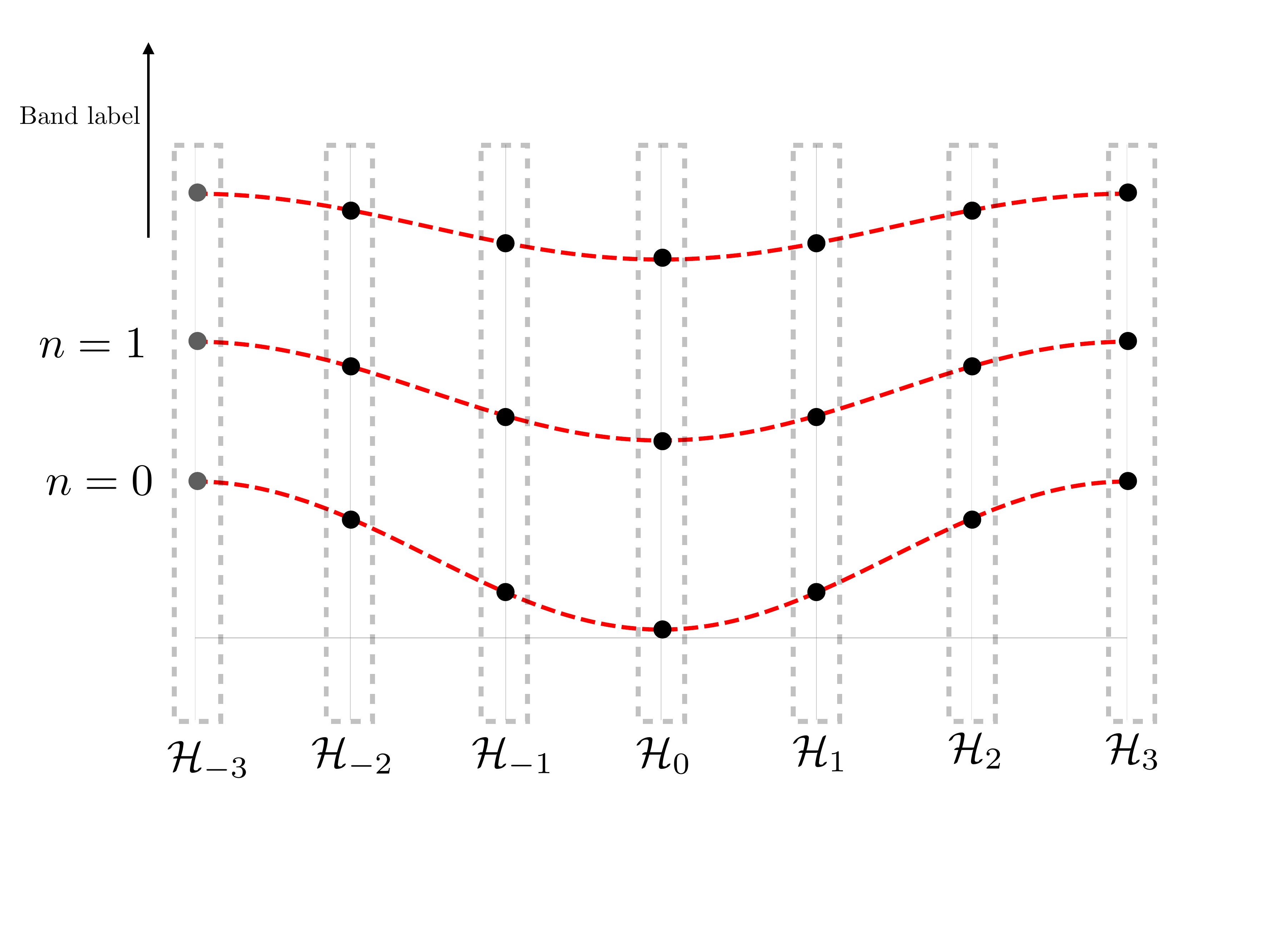}
\vspace{-1.0cm}
\caption{The Hilbert space of the original $T_N$ model can be decomposed according to    $\Z_N$ quantum numbers associated with Bloch momenta $p$.  $\theta_p =  \frac{2 \pi p}{N}$  can be viewed as a   discrete theta angle, or the coefficient of Chern-Simons term in topological gauge theory. 
${\cal H}_p $ acquires an interpretation as Hilbert space of  $(T_N/\Z_N)_p$ theory.  The exact quantization conditions of the $T_N$ theory factorizes into  the quantization conditions for the $(T_N/\Z_N)_p$  models. 
 }
\label{fig:decom}
\end{center}
\end{figure}

We can see the implication of the \eqref{decom} in the partition function. The partition function of the  $T_N$ model  can be written as 
\begin{align}
 Z_0(\beta)  = \tr[e^{-\beta H}] = \sum_{p=0}^{N-1}   \left(  \sum_{n}  e^{-\beta E_{n, p}}  \right)  \equiv   \sum_{p=0}^{N-1}   \widetilde Z_p(\beta)
 \label{pf}
\end{align}
where  $ \widetilde Z_p(\beta) $ is the partition function of the sub-system with  fixed Bloch momentum $p$. 
Let us also define the partition function with the insertion of translation operator, 
\begin{align}
 Z_\ell  (\beta) & =   \tr [e^{-\beta H}  \mathsf {U}^\ell  ]   = \sum_{p=0}^{N-1}  e^{ i 2\pi p  \ell /N}    \tilde Z_p(\beta) 
   \label{tpf}
 \end{align}
This  is just regular partition function for $\ell=0$.  $Z_\ell   (\beta) $ and  $\widetilde Z_p(\beta) $ are related via a discrete Fourier transformation.

Now, we can describe in an elementary way gauging  of   $\Z_N$ symmetry, see   \cite{Kikuchi:2017pcp,Gaiotto:2017yup}.
Physically, gauging translation symmetry  is the declaration that  
$x $  and $x + \frac{2\pi}{N}$ are physically equivalent points.  So, the size of the $S^1$ circle is reduced from $2\pi$ down to     $\frac{2\pi}{N}$.  
Hence, there is only one minimum on the fundamental domain of gauged $T_N/\Z_N$ theory. This means, on each band on the Hilbert space, instead of $ N$  states, only one state survives, i.e.  each band is diluted by a factor of $N$. 
  
  More formal description of gauging is as follows. Global symmetry generators are a set of  codim- 1 defects,  (point defects in the present case). 
  We can gauge the discrete global symmetries by summing over all possible networks of such codim-1 defects, namely, summing over 
  $\ell$,   
  \begin{align}
 \widetilde Z_0(\beta)    \equiv Z_{( T_N/\Z_N)_0}   =  \tr [e^{-\beta H} \Pi_0 ]  & \equiv   \frac{1}{N}  \sum_{\ell=0}^{N-1}    \; Z_\ell   
  \label{gauged-0}
 \end{align}
 The projection to zero Bloch momentum states guarantees this partition function  corresponds to the Hilbert space ${\cal H}_0 $ in the Bloch  decomposition  \eqref{decom}.  
  However, this is not the only  possibility for gauging.  The  gauging procedure admits the freedom to add a topological phase, a discrete topological theta angle, $\theta_p=  \frac{2 \pi  p }{N}$  to each network configuration of the topological defects.  This is equivalent to the insertion of other projection operators \eqref{proj} into the state sum.   Hence, 
  \begin{align}
  \widetilde Z_p(\beta)  \equiv Z_{( T_N/\Z_N)_p}  =   \tr [e^{-\beta H} \Pi_p ]  \equiv   \frac{1}{N}  \sum_{\ell=0}^{N-1} e^{- \im \frac{2 \pi \ell p }{N}}     \; Z_\ell   
  \label{gauged}
 \end{align}
 Using the fact that    $Z_\ell  =   \sum_{n}  \sum_{k =0}^{N-1}   e^{ \im \frac{2 \pi \ell k}{N}}   e^{ -\beta E_{n,k}   }  $ where $n$ is band and $k$ 
 is Bloch momentum label,  we can immediately deduce that the sum reduce to 
 \begin{align}
  \widetilde Z_p(\beta) \equiv Z_{( T_N/\Z_N)_p}   &= \sum_{ |n, p\rangle  \in {\cal H}_p }  e^{  -\beta E_{n,p}    } 
  \label{projection}
 \end{align}
 This  is just the set of states in the  Hilbert space ${\cal H}_p $ in the decomposition  \eqref{decom}.    
 
 As a result of decomposition of Hilbert space according to discrete theta angle $\theta_p$, the quantization condition in the $T_N$  model with topological theta angle  $\theta$  and  
the one in the ${( T_N/\Z_N)_p}$  theories must be related by the factorization formula:
  \begin{align}
%  D_{T_N} (E, \theta)  = \prod_{p=0}^{N-1}   D \left(E,  \frac{\theta + 2 \pi p} {N}   \right) 
 D_{T_N} (E)  = \prod_{p=0}^{N-1}   D_{(T_N/\Z_N)_p} \left(E  \right) 
    \label{relation}
 \end{align}
 in exact agreement with the formula \eqref{eq:Airy_N} obtained from  exact  WKB analysis.

 \subsection{Factorization of exact quantization from Path integral description}
There is also some benefit to  be gained to present this construction in path integral,  especially,  for the identification of  Bloch momentum (which acts as a label in the  decomposition of Hilbert space), which in turn is a discrete theta angle.  The origin of this term is a topological Chern-Simons term in quantum mechanics \cite{Gukov:2013zka, Kapustin:2014gua}.  This discussion is slightly more abstract compared to our explicit Hilbert space construction, but  it generalize more naturally to QFT.   For this reason, we provide a short over-view of the path integral formulation of coupling of QM to $\Z_N$ topological gauge theory.

Since we will ultimately gauge the $\Z_N$   global symmetry \eqref{sym} in our quantum mechanical system,  it is first useful to describe 
 $\Z_N$ topological gauge theory. 
 The $\Z_N$ gauge field can be described  by a pair  of fields  $(A^{(1)}, A^{(0)})$,  which obeys 
\begin{align}
N A^{(1)}= \diff A^{(0)}, \qquad  \int A^{(1)}=  \frac{1}{N}  \int \diff A^{(0)} = \frac{ 2 \pi}{N} \ell, \qquad \ell \in   \Z
\label{ZN}
\end{align} 
proper $\Z_N$ quantization.  The fact that we denoted the holonomy of  $\int A^{(1)}$ field as $\ell$ is not an accident, and is tied with the insertion of 
$\Z_N$ generators $\mathsf U^{\ell}$ into the partition function \eqref{tpf}.    
The partition function of  $\Z_N$ topological gauge theory can be written as 
  \begin{align}
 Z_{{\rm top}, p} &= \int  \Diff   A^{(1)} \Diff A^{(0)}    \Diff  F^{(0)} \;  e^{ \im \int F^{(0)}  \wedge ( N A^{(1)}-\diff A^{(0)}) + \im p \int A^{(1)}}  
 \label{top}
 \end{align}
where $F^{(0)}$ is Lagrange multiplier, which forces \eqref{ZN}, and $i p \int A^{(1)}$  is  the Chern-Simons term in $1d$. 
The action has gauge redundancy   $A^{(1)} \mapsto   A^{(1)}  + \diff \lambda^{(0)},   A^{(0)} \mapsto   A^{(0)} + N \lambda^{(0)} $ and $F^{(0)} \mapsto   F^{(0)} $. 

To couple  dynamical field $x$ to  the $\Z_N$  background  field,  we declare 
 \begin{align}
x \mapsto  x - \lambda^{(0)}.
\end{align}
As a result, the gauge invariant combinations  are  $Nx+   A^{(0)},  \;  \diff x  + A^{(1)}$, and only they  can appear in the Lagrangian with a classical $\Z_N$ background.  Indeed, the insertion of translation generator \eqref{tpf} needs to be identified with  $Z[ (A^{(1)}, A^{(0)})]$ where   background $\Z_N$  field  is given in \eqref{ZN}.  
\begin{align}
  Z[ (A^{(1)}, A^{(0)})]    
%  & =   \cr  
  &=\int_{
 {\rm pbc}
  }   \Diff x \;  
  e^{   -\frac{1}{g} \int d \tau \left( \half (\dot x + A_\tau)^2 - \cos(Nx + A^{(0)}) \right)  + \frac{ \im \theta}{2 \pi} \int (\diff x   + A^{(1)} )   } \delta( N A^{(1)}-\diff A^{(0)})   \qquad  \cr
 & =  \int_{ \tilde x(\beta)= \tilde  x(0) + \frac{2 \pi}{N}\ell }   \Diff  \tilde x \;    e^{   -\frac{1}{g} \int d \tau \left( \half (\dot  {\tilde x})^2 - \cos(N \tilde x) \right)  + \frac{ \im \theta}{2 \pi} \int \diff  \tilde x     }   \cr 
 &  =   \tr [e^{-\beta H}  \mathsf {U}^\ell  ]  =   Z_{\ell}(\beta)
 \end{align}
 where  pbc denotes periodic boundary conditions  $x(\beta)= x(0)$. 
In the second  line, we converted the  background $\Z_N$  gauge field  into a twisted boundary condition for the path integral, by a field redefinition. In the semi-classical description, this guarantees that the leading  saddle configuration that contributes to this sum is a fractional instanton with topological charge $\ell/N$. 

As described above, gauging $\Z_N$ symmetry amounts to summing over all topological gauge theory backgrounds. Moreover, 
we are allowed  to add  a topological phase to each network configuration of the topological defects, which is 1d  Chern-Simons term. 
As a result, 
\begin{align}
  \widetilde Z_p(\beta) \equiv  Z_{( T_N/\Z_N)_p}   &= 
    \int  \Diff A^{(1)}  \Diff A^{(0)}    \;  \delta( N A^{(1)}-\diff A^{(0)})  \;  Z[ (A^{(1)}, A^{(0)})]  \;  e^{ \im p \int A^{(1)}}     \cr
 & = \frac{1}{N}  \sum_{\ell=0}^{N-1} e^{- \im \frac{2 \pi \ell p }{N}}     \; Z_\ell   
  \label{eq:tpf2}
 \end{align}
This corresponds to the path integral formulation for  the theory with the discrete theta angle $\theta_p$ or equivalently, the Hilbert space projected to
${\cal H}_p, \;  p=0, 1, \ldots, N-1$. 

The $p$ label  that permeates the discussion has multiple equivalent and useful interpretations   \cite{Unsal:2012zj, Kikuchi:2017pcp,Gaiotto:2017yup}.
  1) Discrete theta angle $\theta_p$, 2) Level of Chern-Simons term 
3) Decomposition of Hilbert space  using projection operators  $\Pi_p$, where  $p$ is  Bloch momenta.  
The  factorized terms in the exact quantization condition in the    $T_N $ model can be interpreted as the exact quantization  for the $( T_N/\Z_N)_p $  models.

It is important to note that  the local dynamics and saddles in the original $T_N$ model  whose target space is $S^1$  and gauged  $( T_N/\Z_N)_p $  models whose target space is $S^1/\Z_N$      are same.    
An     instanton   with winding number $1$  and action $S_0$ in 
   the $( T_N/\Z_N) $ model is  what we would call  a fractional instanton  with winding number $\frac{1}{N}$ and the same action $S_0$ in the original 
   $ T_N$ model.      In this sense, it is natural that exact quantization conditions and perturbative/non-perturbative relations  are related in a precise sense. One  utility of this perspective is that the mixed anomalies are naturally encoded into exact WKB analysis.  Another utility is the perspective that resurgence is valid within the semi-classical description of each  $ \widetilde Z_p(\beta) \equiv  Z_{( T_N/\Z_N)_p}$, and this implies that resurgence is closed within each fixed topological sector $Z_{\ell}$ as discussed in section \ref{sec:res-tri}.

%%%%%%%%%%%%%%%%%%%%%%%%%%%%%%%%%%%%%%%%%%%%%%%
\section{Analysis of the degenerate Weber-type Stokes graphs}
\label{sec:WB}
%%%%%%%%%%%%%%%%%%%%%%%%%%%%%%%%%%%%%%%%%%%%%%%

%%%%%%%%%%%%%%%%%%%%%%%%%%%%%%%%%%%%%%%%%%%%%%%
\subsection{Relation between Airy-type and degenerate Weber-type Stokes graphs}
%%%%%%%%%%%%%%%%%%%%%%%%%%%%%%%%%%%%%%%%%%%%%%%
In this section, we investigate the quantization condition for the periodic potential (the $S^{1}$ system) by use of the degenerate Weber(DW)-type Stokes graph.
The DW-type Stokes graph should be used under the assumption that the energy spectrum is classically zero ($E_{0}=0$).  Here, $E_0$ is  a control parameter of the Stokes graph, and the picture of the Airy-type Stokes graph is not naively applicable when taking $E_0=0$ where two turning points merge into one. 
Due to the dependence of the Stokes graph on the value of $E_0$, two turning points giving a primary perturbative cycle in the Airy-type Stokes graph collides with each other and merge into a single turning point.
The perturbative cycles are given by a residue integration around the merged turning points consequently, while nonperturbative cycles is defined in the same manner as that of the Airy-type graph.
For the DW-type Stokes graph, one has to employ the connection formula obtained by the DW-type Schr\"{o}dinger equation given by
\be
\left[ - \hbar^2\frac{\pd^2}{\pd y^2} + \frac{y^2}{4} - \hbar \kappa \right] \widehat{\psi}(y,\hbar) = 0,
\label{eq:DWS}
\ee
with $\kappa$ being determined by the global potential.
%{\color{red} Clarify: 
%It is notable that $\hbar$ remains in the equation for this case.
%We also emphasize that the resulting  quantization condition directly gives the quasi-moduli contributions (bion contributions). Also, let us change hat symbol to widehat... The latter is more easy to see.}

Let us address the relation between the quantization conditions based on the Airy-type and the DW-type Stokes graphs.
By rescaling the energy as $E\rightarrow \hbar E$ in the Schr\"{o}dinger equation, the Stokes graph of the system transforms from the Airy-type to the DW-type as discussed in \cite{ZinnJustin:2004cg,DP1}, and the monodromy matrix in the DW-type leads to a different connection formula from the Airy-type. 
The derivation of the connection formula of the DW-type is given  in Appendix~\ref{sec:der_DWconn}, which is obtained by starting with the DW equation and performing the coordinate transformation.

The connection formula in the Airy- and DW-type seem to be different from each other, but the most important fact we show in this section is that there is a \textit{dictionary} to translate expressions of cycles from one into the other, which we derive in Appendix~\ref{sec:dic_A_DW}.
Owing to the dictionary, one can immediately find, not only that the DDP formula and the resurgent structure are unchanged, but also that the techniques such as the resolvent method and the Gutzwiller formula can be performed just by  the substitution: 
\begin{align}
  {\rm Airy:}\;\; (A,B) \longrightarrow {\rm Degenerate \;  Weber:}\;\;   ({\cal A},{\cal B})
\end{align}
and following the dictionary shown in Table.~\ref{eq:Airy-Weber_dictionary}.
%symbols $(A,B)$ with $({\cal A},{\cal B})$ by following the dictionary shown in Table.~\ref{eq:Airy-Weber_dictionary}.

Although the two expressions of $D(E)$ derived from Airy-type and DW-type  Stokes graphs  give equivalent Fredholm determinants,   they have their own advantages and disadvantages due to simple fact that different information are encoded in  different places.  The former gives a clear Gutzwiller representation and makes it easier to see the relationship between the respective quantization methods, but to solve $D(E)=0$ for $E$ accurately is difficult. The latter gives an accurate expression of quasi-moduli-integral  and bion amplitude  
because the construction  zoom into ($E_{0}=0$) accurately, but the connection formula becomes relatively complicated. 
% by solving $D(E)=0$  in the path integral representation of the partition function, but the connection formula becomes complicated. 

In the next subsection, we discuss  %the partition function and the resurgent structure from  
the quantization condition by the exact-WKB analysis with the degenerate-Weber-type(DW-type) Stokes graph.

\begin{figure}[t]
    \centering 
    \includegraphics[width=8cm]{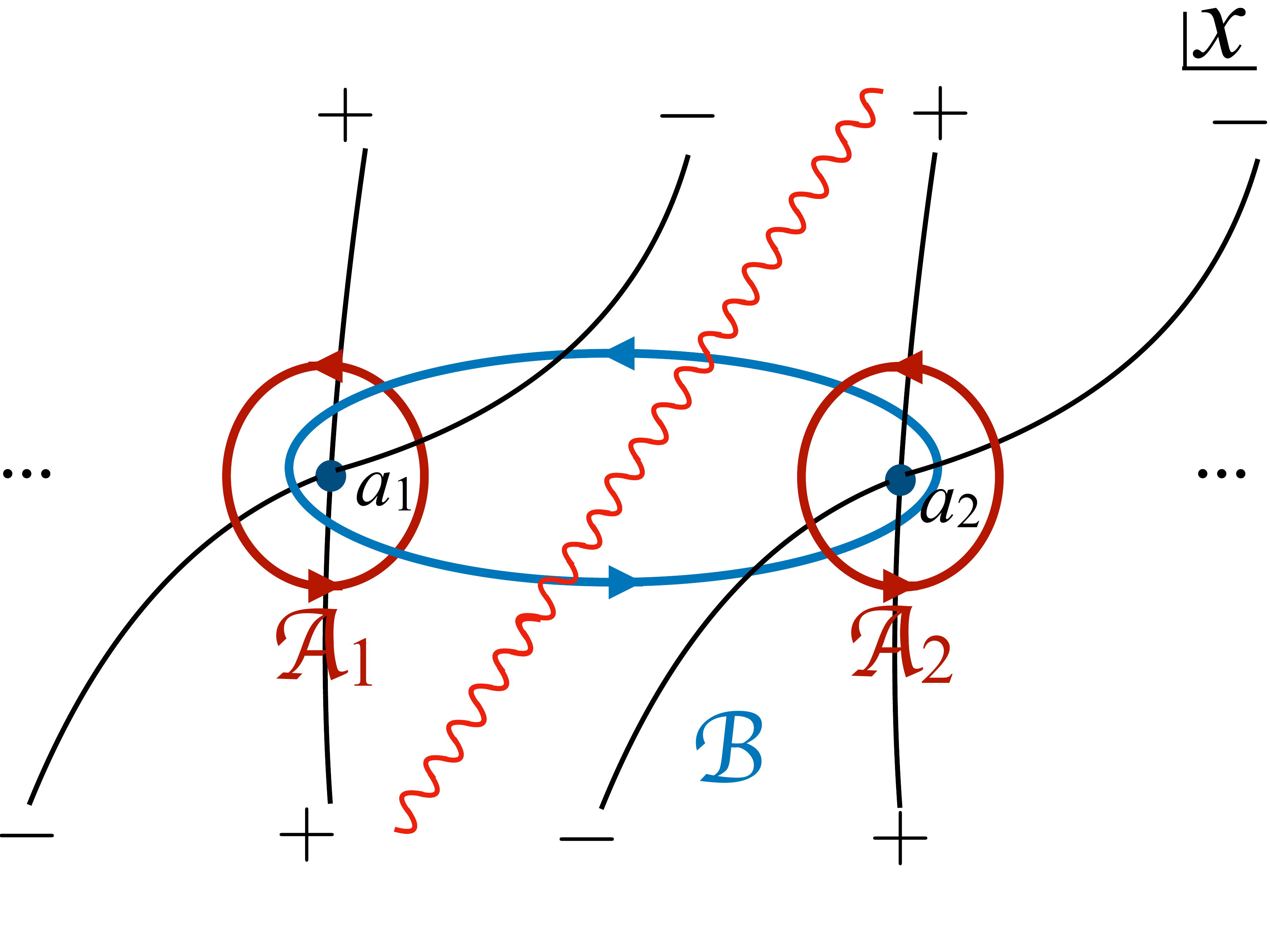}
    \caption{The local degenerate Weber-type Stokes graph is depicted. We here consider two different ${\cal A}$ cycles for generality, but they are identical for our present system with periodic potential.}
     \label{fig:S1-weber}
\end{figure}

%%%%%%%%%%%%%%%%%%%%%%%%%%%%%%%%%%%%%%%%%%%%%%%
%\subsection{From quantization condition to trans-series representation}
\subsection{From quantization condition to partition function}
\label{Sec:Partition}
%%%%%%%%%%%%%%%%%%%%%%%%%%%%%%%%%%%%%%%%%%%%%%%

\begin{figure}[t]
  \begin{center}
    \begin{tabular}{cc}
      \begin{minipage}{0.4\hsize}
        \begin{center}
          \includegraphics[clip, width=70mm]{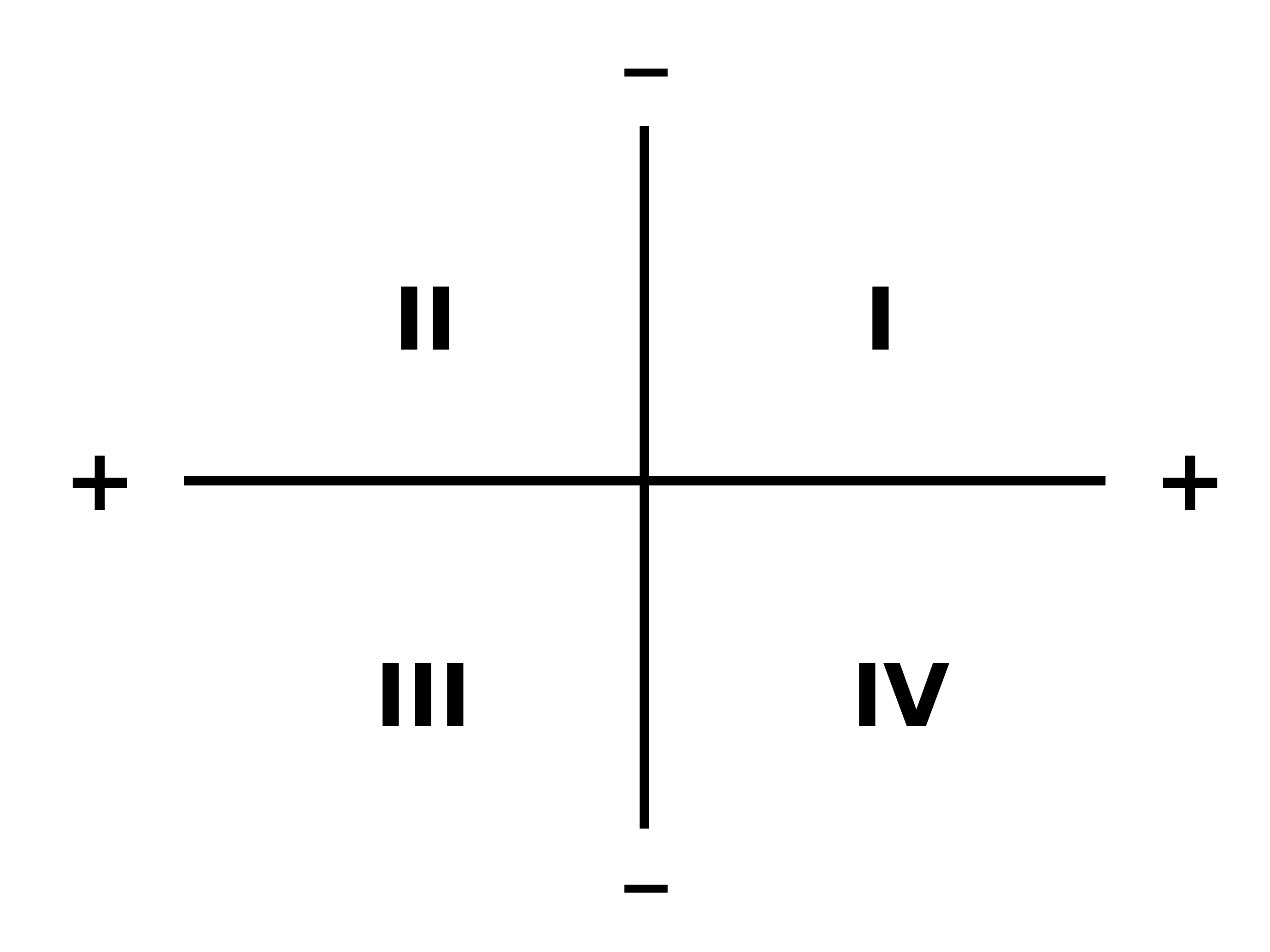}
          \hspace{1.6cm} 
        \end{center}
      \end{minipage}   \quad 
      \begin{minipage}{0.4\hsize}
        \begin{center}
          \includegraphics[clip, width=70mm]{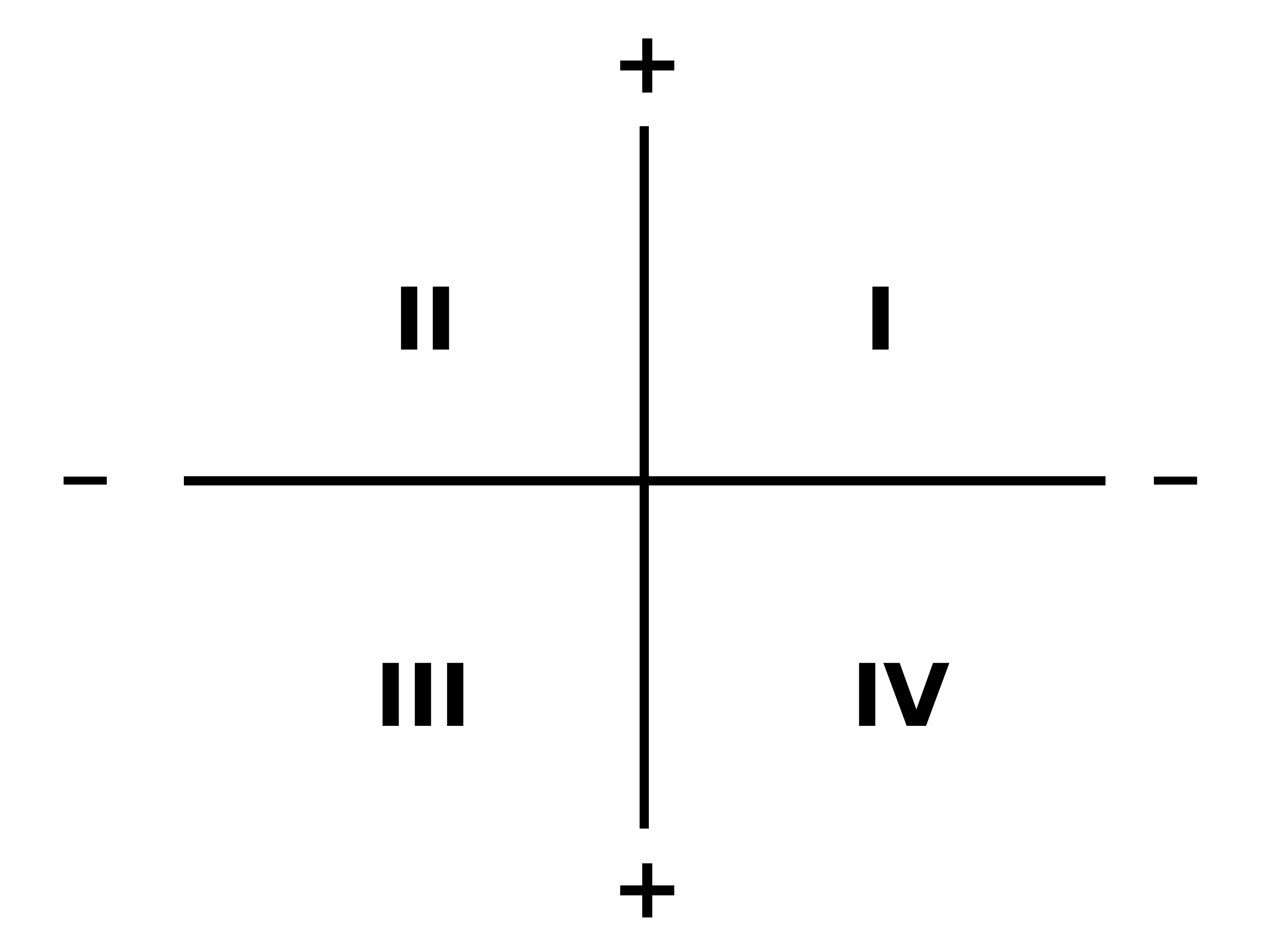}
          \hspace{1.6cm} 
        \end{center}
      \end{minipage}      
    \end{tabular} 
    \caption{The Stokes curve for the degenerate Weber equation. $\pm$ labels the asymptotics of $\widehat{S}_{\rm odd,-1}(y)$.}
    \label{fig:deg_web_loc}
  \end{center}
\end{figure}

Let us begin with showing the connection formula of the DW Stokes graph.
We now consider the local Stokes graph as shown in Fig.~\ref{fig:S1-weber}, where the cycles are denoted as ${\cal A}$ and ${\cal B}$.
For generality, we denote two different ${\cal A}$ as ${\cal A}_{1}$ and ${\cal A}_{2}$, but they are identical in our present problem with the periodic potential.

From now, we consider the potential $V(x)=1-\cos (Nx)$, which has ${\mathbb Z}_{N}$ symmetry. 
In the DW type Stokes graph, the region around a turning point is separated to four regions, that we denote as  I, II, III and IV, as shown in Fig.\ref{fig:deg_web_loc}. %of  Appendix \ref{sec:dweq}. 
This is unlike Airy type graphs which splits to three regions. 
The monodromy matrices are given by \cite{DDP2,Takei3}
\be
&&   {\cal M}^{\rm IV \rightarrow I}_{\ominus} =
   \begin{pmatrix}
     1 &  0 \\
     i \frac{C_{-}(E,\hbar)}{C_{+}(E,\hbar)}\frac{\sqrt{2 \pi} \, e^{+\pi i F(E,\hbar)}\, \hbar^{+F(E,\hbar)}}{\Gamma(1/2 - F(E,\hbar))}   & 1 
   \end{pmatrix}, \quad
   {\cal M}^{\rm I \rightarrow II}_{\oplus} =
   \begin{pmatrix}
     1 &  i \frac{C_{+}(E,\hbar)}{C_{-}(E,\hbar)}\frac{\sqrt{2 \pi} \, \hbar^{-F(E,\hbar)}}{\Gamma(1/2 + F(E,\hbar))} \\
     0 & 1 
   \end{pmatrix}, \nl
&&   {\cal M}^{\rm II \rightarrow III}_{\ominus} =
   \begin{pmatrix}
     1 &  0 \\
     i \frac{C_{-}(E,\hbar)}{C_{+}(E,\hbar)}\frac{\sqrt{2 \pi}\, e^{-\pi i F(E,\hbar)}\, \hbar^{+F(E,\hbar)}}{\Gamma(1/2 - F(E,\hbar))}   & 1 
   \end{pmatrix}, \quad
   {\cal M}^{\rm III \rightarrow IV}_{\oplus} =
   \begin{pmatrix} 
     1 &  i \frac{C_{+}(E,\hbar)}{C_{-}(E,\hbar)}\frac{\sqrt{2 \pi} \,e^{-2\pi i F(E,\hbar)}\, \hbar^{-F(E,\hbar)}}{\Gamma(1/2 + F(E,\hbar))} \\
     0 & 1 
   \end{pmatrix}, \nl   
\ee
%The regions I, II, III and IV stand for the four local regions around the turning point, which we also discuss in Fig.\ref{fig:deg_web_loc} of  Appendix \ref{sec:dweq} in detail. 
The derivation of monodromy matrices is given in Appendix \ref{sec:dweq}. 
Here, $F(E,\hbar)$ are $C_{\pm}(E,\hbar)$ are defined as
\be
&& F(E,\hbar) := {\rm Res}_{x=0} \, S^{\rm DW}_{\rm odd}(x,E,\hbar) \approx -\frac{E}{N}, \label{eq:def_F} \\
&& C_{\pm}(E,\hbar) := \lim_{x \rightarrow 0}\left( \frac{\pd y(x,E,\hbar)}{\pd x} \right)^{1/2} \frac{{\psi}_{\pm}(x,E,\hbar)}{\widehat{\psi}_{\pm}(y(x,E,\hbar),E,\hbar)}
%:= \exp \left[\pm \frac{F(\hbar)}{2} \log 2 \right] 
\approx \left( \frac{32}{N} \right)^{\mp \frac{E}{2N}}, 
\label{eq:Cpm}
\ee
respectively, where $a_\ell$ is a turning point connecting with its Stokes line.
 $C_{\ell\pm}(E,\hbar)$ comes from the local coordinate transformation from the DW-type Schr\"{o}dinger equation in Eq.~(\ref{eq:DWS}).
$F(E,\hbar)$ is directly related with the ``quantum'' frequency as $\omega_{\cal A}(E,\hbar)=-E/F(E,\hbar)$ and $\omega_{\cal A}(E,\hbar) = N + O(\hbar)$ in the present system.
We also define the normalization matrix (Voros multiplier, which accounts   the  change  of  turning  points in the   WKB  wave-function)  and the branch-cut matrix as
\be
&& {\cal N}_{a_1, a_2} := 
   \begin{pmatrix}
     e^{+\frac{1}{\hbar} \int^{a_2}_{a_1} dx \, S^{\rm DW}_{\rm odd,-1}(x)}  & & 0\\
     0 & & e^{-\frac{1}{\hbar} \int^{a_2}_{a_1} dx \, S^{\rm DW}_{\rm odd,-1}(x)}
   \end{pmatrix}, \\
&&   {\cal T} := 
   \begin{pmatrix}
     0 & & -i \\
     -i & & 0
   \end{pmatrix},
\ee
respectively, where $a_1$ and $a_2$ are turning points as $a_2 = a_1 + 2 \pi \sim a_1$.
%The details of derivation for these monodromy matrices are shown in Appendix~\ref{sec:der_DWconn}.

The quantization condition for the $T_N$ model with $\Z_N$ global symmetry is expressed as 
\be
{\cal D}^{(N)\pm} =  {\rm det} \left[ \left( {\cal M}^{\pm} \right)^N - I e^{i \theta} \right] = 0,
\ee
where $\theta$ is the contribution from the $\theta$-angle (or the twisted boundary condition applied to WKB wave-function), and
\be
&& {\cal M}^{+} = {\cal M}_{\oplus}^{{\rm III} \rightarrow {\rm IV}} \, {\cal N}_{a_1,a_{2}} \,  {\cal T} \, {\cal M}_{\ominus}^{{\rm II} \rightarrow {\rm III}}, \qquad {\cal M}^{-} = {\cal M}_{\oplus}^{{\rm III} \rightarrow {\rm IV}} \, {\cal M}_{\ominus}^{{\rm IV} \rightarrow {\rm I}}  \, {\cal N}_{a_1,a_{2}} \, {\cal T}.
\ee
Thus, ${\cal D}^{(N)\pm}$ can be expressed as
\be 
  {\cal D}^{(N)\pm} 
&=& \frac{1}{({\cal A}^{\mp 1} {\cal B})^{N/2}} \prod_{p=0}^{N-1} \left[ 1 +{\cal A}^{\mp 1} \left(1 +  {\cal B} \right) - 2 \sqrt{{\cal A}^{\mp 1}{\cal B}} \cos \left( \frac{ \theta + 2 \pi p}{N}\right) \right]. \label{eq:DNN} 
\ee
Here, the symbolic notation of cycles, ${\cal A}$ and ${\cal B}$, are expressed as 
\be
&& {\cal A} := e^{2 \pi i F} \approx e^{-2 \pi i E/N}, \label{eq:cal_A} \\
&& {\cal B} := \left(\frac{C_-}{C_+}\right)^2 \frac{2 \pi  {\cal B}_0 \hbar^{2F}}{\Gamma \left(\frac{1}{2}-F \right)^2} \approx \frac{2 \pi  {\cal B}_0}{\Gamma \left(\frac{1}{2}+\frac{E}{N} \right)^2} \left( \frac{N\hbar}{32} \right)^{-\frac{2E}{N}}, \label{eq:cal_B}
%{\cal B}_0 %=e^{-\frac{1}{\hbar} \oint_{\pi}^\pi dx \, \sqrt{Q_0(x)}}
% = e^{-\frac{16}{\hbar}}.
\ee
where ${\cal B}_0 = e^{-\frac{16}{N \hbar}}$ denotes the exponential of (minus) action of the  bion.  
%and 
%\be
%&& {\cal A}(\hbar) = e^{- 2 \pi i \frac{E}{\omega_{\cal A}(E,\hbar)}}, \nl
%&& F(E,\hbar) = -\frac{E}{\omega_{\cal A}(E,\hbar)}, \nl
%&& {\cal B}(\hbar)
%= \frac{2 \pi  {\cal B}_0 
%M_{\cal B}(\frac{E}{\omega_{\cal A}(E,\hbar)},\hbar)}{\Gamma (\frac{1}{2}+\frac{E}{\omega_{\cal A}(E,\hbar)})^2} \left( \frac{\hbar}{2} \right)^{-2E/\omega_{\cal A}(E,\hbar)}.
%\label{eq:Airy-Weber_dictionary}
%\ee

\begin{table}[t]
  \begin{tabular}{|c|c|r||r|} \hline
    Airy-type & degenerate Weber-type\\ \hline 
    & \\ 
    $A_{\ell}=e^{\oint_{A_\ell}dx \, S_{\rm odd}}$ & ${\cal A}_\ell=e^{-\frac{2\pi i E}{\omega_{{\cal A}_{\ell}(E,\hbar)}}} \approx e^{-\frac{2 \pi i E}{\omega_{{\cal A}_{\ell}}}}$  \\
    & \\ \hline & \\
    $B=e^{\oint_{B}dx \, S_{\rm odd}}$ & 
    \begin{tabular}{c}
    $\mathcal{B}=2 \pi  e^{-\frac{S_B}{\hbar}} \displaystyle{\prod_{\ell=1}^2} \frac{C_{\ell-}(E,\hbar)}{C_{\ell+}(E,\hbar)} 
    \frac{e^{(-1)^{\ell}\frac{\pi i E}{\omega_{{\cal A}_\ell}(E,\hbar)}}
    \hbar^{-\frac{E}{\mathcal{\omega_{A_\ell}}(E,\hbar)}}}{\Gamma (\frac{1}{2}+\frac{E}{\mathcal{\omega_{A_\ell}}(E,\hbar)})}$\\
    $\approx 2 \pi  e^{-\frac{S_B}{\hbar}} \displaystyle{\prod_{\ell=1}^2} \frac{e^{(-1)^{\ell}\frac{\pi i E}{\mathcal{\omega_{A_\ell}}}}}{\Gamma(\frac{1}{2}+\frac{E}{\mathcal{\omega_{A_\ell}}})} \left( \frac{N\hbar}{32}\right)^{-\frac{E}{\omega_{{\cal A}_\ell}}}$
    \end{tabular}\\ & \\ \hline  
  \end{tabular}
  \caption{Dictionary of $A$ and $B$ cycles between the Airy-type and the DW type. The quantum frequency $\omega_{\cal A}(E,\hbar)$ has the relationship with $F(E,\hbar)$ in Eq.(\ref{eq:def_F}) as $\omega_{\cal A}(E,\hbar) = -E/F(E,\hbar) \approx \omega_{\cal A} + O(\hbar)$. Furthermore, $S_B=2\int_{a_1}^{a_2}\sqrt{2V(x)}dx \in {\mathbb R}_{>0}$ is the classical bion action. For the periodic potential, all of $A$-cycles are identical.}
  \label{eq:Airy-Weber_dictionary}
\end{table}

Now, we discuss the relation between the Airy- and the DW-type quantization conditions.
Comparing Airy-type and DW type quantization conditions, one finds the \textit{dictionary} in Tab.~\ref{eq:Airy-Weber_dictionary}, where $D^\pm(E)$ obtained from Airy-type and $\mathcal{D}^\pm(E)$ obtained from DW-type are symbolically identical. In other words, if we apply the dictionary in Tab.~\ref{eq:Airy-Weber_dictionary} to $D(E)^\pm$ obtained from Airy-type, we easily obtain the Weber-type one.
The derivation of the dictionary is reviewed in Appendix~ \ref{sec:dic_A_DW}. 
This  dictionary is   applicable to the generic one-dimensional potentials, not only  periodic ones. 

We here investigate the DW type quantization condition in Eq.~(\ref{eq:DNN}) in detail.
We first derive the nonperturbative contribution to the ground state energy.
Based on the approximations in Eqs.(\ref{eq:cal_A}) and (\ref{eq:cal_B}), the quantization condition for $N=1,2$ are expressed as
\be
   {\cal D}^{(1)\pm}(E)
   &\approx&    \frac{1}{\sqrt{{\cal B}_0}\Gamma\left(\frac{1}{2}-E\right)} \left( \frac{\hbar}{32} \right)^{E}+ \frac{\sqrt{{\cal B}_0} e^{\pm \pi i E}}{\Gamma \left(\frac{1}{2}+E \right)} \left( \frac{\hbar}{32} \right)^{-E} - \sqrt{\frac{2}{\pi}}  \cos \theta = 0, \label{eq:DN1_lowE} \\
   {\cal D}^{(2)\pm}(E)
&\approx& \left[ \frac{1}{\sqrt{{\cal B}_0} \Gamma \left(\frac{1}{2}-\frac{E}{2} \right)}   \left( \frac{\hbar}{16} \right)^{\frac{E}{2}}  + \frac{\sqrt{{\cal B}_0}e^{\pm \pi i \frac{E}{2}}}{\Gamma \left(\frac{1}{2}+\frac{E}{2} \right)} \left( \frac{\hbar}{16} \right)^{-\frac{E}{2}} + \sqrt{\frac{2}{\pi}} \cos \frac{\theta}{2}  \right] \nl
   && \cdot \left[ \frac{1}{\sqrt{{\cal B}_0} \Gamma \left(\frac{1}{2}-\frac{E}{2} \right)}   \left( \frac{\hbar}{16} \right)^{\frac{E}{2}}  + \frac{\sqrt{{\cal B}_0}e^{\pm \pi i \frac{E}{2}}}{\Gamma \left(\frac{1}{2}+\frac{E}{2} \right)} \left( \frac{\hbar}{16} \right)^{-\frac{E}{2}}  - \sqrt{\frac{2}{\pi}} \cos \frac{\theta}{2}  \right] = 0, 
\ee
up to an irrelevant overall factor. % being irrelevant to the partition function.
Now, we would set the energy as $E/N=1/2+\delta$ with $0 < |\delta| \ll 1$.
Solving the quantization condition in terms of $\delta$ yields
\be
\delta^{N=1} &=& - \sqrt{\frac{64{\cal B}_0}{\pi \hbar}} \cos \theta  +  \frac{64{\cal B}_0  }{\pi \hbar } \left[ \cos^2 \theta \cdot \left(  \gamma -\log \frac{ \hbar }{32} \right) \pm \frac{\pi i}{2}  \right] + O({\cal B}^{3/2}_0) \\
\delta^{N=2} &=& - (-1)^p \sqrt{\frac{32{\cal B}_0}{\pi \hbar}} \cos \frac{\theta}{2}  +  \frac{32{\cal B}_0  }{\pi \hbar } \left[ \cos^2 \frac{\theta}{2} \cdot \left(  \gamma -\log \frac{ \hbar }{16} \right) \pm \frac{\pi i}{2} \right] + O({\cal B}^{3/2}_0), \label{eq:deltaN2}
%\delta^{N=1} &=& -  \frac{8{\cal B}_0^{1/2}}{\sqrt{\pi} \hbar^{1/2}} \cos \theta \pm  i  \frac{32{\cal B}_0 }{\hbar}  + \frac{256{\cal B}_0^{3/2} \cos \theta}{\sqrt{\pi} \hbar ^{3/2}}  \left[\pi \mp i \left( \gamma -\log  \frac{\hbar}{32} \right) \right] \nl 
%&& -\frac{1024 {\cal B}_0^2}{\hbar^2} \left[ \gamma -\log  \frac{\hbar}{32}  \pm i \pi   \right] +O\left({\cal B}_0^{5/2}\right), \\
%\qquad  \, \mbox{for $N=1$}, \\ 
%\delta^{N=2} &=& (-1)^p \frac{4 \sqrt{2}{\cal B}_0^{1/2} \cos \frac{\theta}{2}}{\sqrt{\pi} \hbar^{1/2}} \pm  i  \frac{16{\cal B}_0 }{\hbar} 
%+ (-1)^p \frac{64 \sqrt{2} {\cal B}_0^{3/2} \cos \frac{\theta}{2}}{\sqrt{\pi} \hbar ^{3/2}} \left[\pi \mp i \left( \gamma -\log  \frac{\hbar}{16} \right) \right]  \nl 
%&& -\frac{256 {\cal B}_0^2 }{\hbar ^2} \left[ \gamma -\log  \frac{\hbar}{16}  \pm i \pi \right] + O\left({\cal B}_0^{5/2}\right), \label{eq:deltaN2}
%\qquad \, \mbox{for $N=2$},
\ee
where $\gamma$ is the Euler constant, and $p \in \{ 0,+1 \}$ is the eigenvalue of the  ${\mathbb Z}_2$-shift symmetry generator. % given by $x \rightarrow x+p \pi$.
For $N=1$, the first term is the contribution of instanton $[I]$ and anti-instanton $[\bar I]$. The second term can be viewed as  the contributions of correlated events, 
$[II]$,  $[\bar I \bar I]$, and $[I \bar I]_{\pm}$. The imaginary ambiguity originates in $O({\cal B}_0)$ and is sourced by  the $[I \bar I]_{\pm}$ critical point at infinity. 
This contribution is  cancelled  with the same ambiguity that arise from the lateral Borel resummation of perturbation theory. the quantization condition for $N=1$ in the low energy limit in Eq.(\ref{eq:DN1_lowE}) is in exact agreement with the result by Zinn-Justin and Jentsuchura \cite{ZinnJustin:2004cg} up to rescaling parameters in the theory.

%Notice that the imaginary ambiguity originates in $O({\cal B}_0)$ and is purely imaginary, so that this contribution would be cancelled after taking the Borel resummation.
For $N=2$, there are two types of  fractional instantons,  $[I_1]$,    $[I_2]$ and $[\bar I_1]$,    $[\bar I_2]$, 
and at second order, there are $[I_1I_2]$, $[\bar I_1 \bar I_2]$,  $[I_1 \bar I_1]_{\pm}$,  $[I_2 \bar I_2]_{\pm}$ 
critical points at infinity, whose manifestation in \eqref{eq:deltaN2} is clear. 
In addition to these similar  effects with the  $N=1$ case,  there is one more interesting phenomenon.    At $\theta=\pi$,  degenerate energies independent of $p$ is obtained.  In fact, the whole Hilbert space is two-fold degenerate.  As discussed in the previous section, this is a manifestation of mixed anomaly between the two global symmetries of the theory,  $\Z_2 \times \Z_2$, translation and charge conjugation, and anomaly implies the spontaneous breaking down to $\Z_2$ leading to two-vacua. 

%For this case the single-instanton effect to the energy vanishes, and the real part of nonperturvative part starts with $O({\cal B}_0^{2})$.It is important to mention that the quantization condition for $N=1$ in the low energy limit in Eq.(\ref{eq:DN1_lowE}) is in exact agreement with the result by Zinn-Justin and Jentsuchura \cite{ZinnJustin:2004cg} up to rescaling parameters in the theory.%\footnote{
%In \cite{ZinnJustin:2004cg}, the authors were beginning with $V(x)=\frac{1}{8}[1-\cos (4 x)]$.
%}

Let us now express  the partition function by use of the resolvent method. We restrict to  $N=1$ case for simplicity. 
We  separate the partition function into the perturbative and nonperturbative parts as 
\be
Z(\hbar,\beta) = Z_{\rm pt}(\hbar,\beta) + Z_{\rm np}(\hbar,\beta).
\ee
Since one finds that
\be
   {\cal D}^{(1)\pm} &\propto& 1 + {\cal A}^{\mp 1} \left[ 1 + {\cal B} \right] - 2 \sqrt{{\cal A}^{\mp 1} {\cal B}} \cos \theta \nl
%   &=& 1 + {\cal A}^{\mp 1} + {\cal A}^{\mp 1}   {\cal B}  - 2 \sqrt{{\cal A}}^{\mp 1} \sqrt{{\cal B}} \cos \theta \nl
   &=& {\cal D}_{\cal A}^{\mp} \left[ 1 + \frac{{\cal A}^{\mp 1}}{{\cal D}_{\cal A}^{\mp}}   {\cal B}  -  \frac{2\sqrt{{\cal A}}^{\mp 1}}{{\cal D}_{\cal A}^{\mp}} \sqrt{{\cal B}} \cos \theta \right],
\ee
where
\be
   {\cal D}_{\cal A}^{\pm}(E,\hbar)   = 1 + {\cal A}(E,\hbar)^{\pm 1} 
   = 1 + e^{\mp 2\pi i\frac{E}{\omega_{{\cal A}}(E,\hbar)}}
   =  \frac{2 \pi e^{\mp \pi i \frac{E}{\omega_{\cal A}(E,\hbar)}}}{\Gamma (\frac{1}{2}+\frac{E}{\omega_{\cal A}(E,\hbar)}) \Gamma (\frac{1}{2}-\frac{E}{\omega_{\cal A}(E,\hbar)})}, 
\ee
The perturbative part can be found easily as
\be
Z^{N=1}_{\rm pt}(\hbar,\beta) = \frac{1}{2 \pi i} \int^{\epsilon + i \infty}_{\epsilon - i \infty} \left[ - \frac{\pd \log {\cal D}_{\cal A}^{\mp}}{\pd E} \right] e^{-\beta E} dE,
\ee
and the nonperturbative part can be written as the expanded form in terms of the instanton and bion contributions as:
\be
Z^{N=1}_{\rm np}(\hbar,\beta) &=& \frac{1}{2 \pi i} \int^{\epsilon + i \infty}_{\epsilon - i \infty} \left[ - \frac{\pd}{\pd E} \log\left( 1 + \frac{{\cal A}^{\mp 1}}{{\cal D}_{\cal A}^{\mp}}   {\cal B}  -  \frac{2\sqrt{{\cal A}}^{\mp 1}}{{\cal D}_{\cal A}^{\mp}} \sqrt{{\cal B}} \cos \theta \right)  \right] e^{-\beta E} dE \nl
&=&  \frac{\beta}{2 \pi i} \int^{\epsilon + i \infty}_{\epsilon - i \infty}  \sum_{n=1}^{\infty} \sum_{m=0}^{n}
\frac{1}{n}
\begin{pmatrix}
  n \\
  m
\end{pmatrix}
\left(- \frac{{\cal A}^{\mp 1}}{{\cal D}_{\cal A}^{\mp}}   {\cal B} \right)^m \left( \frac{\sqrt{{\cal A}}^{\mp 1}}{{\cal D}_{\cal A}^{\mp}} \sqrt{{\cal B}} \qty(e^{i\theta}+e^{-i\theta}) \right)^{n-m} e^{-\beta E} dE \nl
&=& \sum_{\substack{(Q,K) \in {\mathbb Z} \otimes {\mathbb N}_0 \\ |Q|+K>0}} Z^{N=1}_{\rm np}(\hbar,\beta;\{Q,K\}), \\
Z^{N=1}_{\rm np}(\hbar,\beta;\{Q,K\}) &:=&
\frac{\beta}{2 \pi i} \int^{\epsilon + i \infty}_{\epsilon - i \infty}
\frac{1}{|Q|+K}
\begin{pmatrix}
  |Q|+K \\
  K
\end{pmatrix}
 \left( \frac{{\cal B}}{{\cal K}^2} \right)^{ |Q|/2+K}    \nl
&& \cdot \, _2F_1\left(1-K,-K;|Q|+1;-{\cal A}^{\pm 1} \right)  \left( - {\cal A}^{\mp 1} \right)^K  e^{-\beta E + i Q \theta}  dE,
\ee
where $Q$ and $K$ are the topological charge and the number of bions, respectively, and
\be
   {\cal K}: = \sqrt{{\cal A}}^{+1} + \sqrt{{\cal A}}^{-1} %= \frac{{\cal D}_{\cal A}^{\pm}}{\sqrt{{\cal A}}^{\pm 1}}
   = {\cal D}_{\cal A}^{\pm}\sqrt{{\cal A}}^{\mp 1}.
\ee
Using Tab.\ref{eq:Airy-Weber_dictionary}, we can get the quasi-moduli-integral (QMI) form:
\begin{align}
    Z_{\rm np}^{N=1}(\hbar,\beta;\{Q,K\}) =&\frac{\beta}{2 \pi i} \int^{\epsilon + i \infty}_{\epsilon - i \infty}
  \frac{(-1)^K}{|Q|+K}
\begin{pmatrix}
  |Q|+K \\
  K
\end{pmatrix}
 \left[ \frac{e^{-\frac{S_{ B}}{\hbar}}}{2\pi}\Gamma\qty(\frac{1}{2}-\frac{E}{\omega_{\cal A}})^2\qty(\frac{\hbar}{32})^{-\frac{2E}{\omega_{\cal A}}} \right]^{ |Q|/2+K}\nl
&  \cdot \, _2F_1\left(1-K,-K;|Q|+1;-e^{\mp 2\pi i\frac{E}{\omega_{\cal A}}} \right)  \left(e^{\pm 2\pi i\frac{E}{\omega_{\cal A}}} \right)^K  e^{-\beta E + i Q \theta}  dE.
\end{align}
The physical meaning of each term is as follows:
$\beta$ is the exact zero mode of the bion and instanton, $(-1)^K$ is the Maslov index, $\mqty(|Q|+K \\ K)$ and $\frac{1}{|Q|+K}$ are combination and cyclic permutation of $K$-bions and $Q$-instantons, $\qty(\frac{e^{-\frac{S_{B}}{\hbar}}}{2\pi}\Gamma\qty(\frac{1}{2}-\frac{E}{\omega})^2\qty(\frac{\hbar}{32})^{-\frac{2E}{\omega}} )^{|Q|/2+K}$ is nothing but the bion (and instanton) amplitude and QMI integral.
Notice that the label of phase ambiguity is $K$, which is the number of (neutral) bions, not the instanton.
\begin{comment}
\be
Z^{N=1}_{\rm np}(\hbar,\beta)  
&=& \frac{\beta}{2 \pi i} \int^{\epsilon + i \infty}_{\epsilon - i \infty}   \left[  \sum_{n=1}^{\infty} \sum_{m=0}^{n} \frac{1}{n}
\begin{pmatrix}
  n \\
  m
\end{pmatrix}  
\left(   \Gamma(-s)^2   (e^{\pm 2\pi i s}-1)
\frac{{\cal B}_0 
%M_{\cal B}(1/2+s,\hbar)
}{2\pi} \left( \frac{\hbar}{2} \right)^{-1-2s} \right)^{m} \right. \nl
&& \left. \cdot \left( \Gamma(-s) \sqrt{\frac{2{\cal B}_0 %M_{\cal B}(1/2+s,\hbar)
}{\pi}} \left( \frac{\hbar}{2} \right)^{-\frac{1}{2}-s} \cos \frac{\theta}{2} \right)^{n-m} \right]
e^{-\beta \omega_{\cal A}(\frac{1}{2}+s)} \omega_{\cal A} ds. 
\ee
\end{comment}

For generic $N$, the partition function can be written in the way parallel to the case of $N=1$.
Since both of perturbative and nonperturbative contributions is $N$-times of that for the single-periodic case, the nonperturbative part can be written as
\be
Z_{\rm np}(\hbar,\beta) 
&=& \sum_{p=0}^{N-1} \sum_{\substack{(Q_p,K_p) \in {\mathbb Z} \otimes {\mathbb N}_0 \\ |Q_p|+K_p>0}} Z_{\rm np}(\hbar,\beta;\{p,Q_p,K_p \}) \label{eq:Znp} \\
&=& N \sum_{\substack{(Q,K) \in {\mathbb Z} \otimes {\mathbb N}_0 \\ |Q|+K>0}} Z_{\rm np}(\hbar,\beta;\{0,N Q,K \}),\label{eq:Znp2} \\
Z_{\rm np}(\hbar,\beta;\{p,Q_p,K_p \}) &:=&
\frac{\beta}{2 \pi i} \int^{\epsilon + i \infty}_{\epsilon - i \infty}
\frac{e^{2 \pi i p Q_p/N}}{|Q_p|+K_p}
\begin{pmatrix}
  |Q_p|+K_p \\
  K_p
\end{pmatrix}
 \left( \frac{{\cal B}}{{\cal K}^2} \right)^{ |Q_p|/2+K_p}    \nl
&& \cdot \, _2F_1\left(1-K_p,-K_p;|Q_p|+1;-{\cal A}^{\pm 1} \right)  \left( - {\cal A}^{\mp 1} \right)^{K_p}  e^{-\beta E + i Q_p \theta/N}  dE. \label{eq:ZpQK}
\ee
Here, in order to derive Eq.(\ref{eq:Znp2}), we have performed the discrete Fourier transform in Eq.(\ref{eq:Znp}),\footnote{
The subscript $p$ in $Q_p$ and $K_p$ is a dummy index.
}
and consequently the contribution from $Q_p  \in N {\mathbb Z}$ remains.
This fact is directly seen in the partition function given through the resolvent method.
For example, by employing the result in Eq.(\ref{eq:deltaN2}), the partition function for $N=2$ with the nonperturbative contribution can be approximately estimated by
\be
\left. Z^{N=2}(\hbar,\beta) \right|_{\rm ground \ state}  &\approx& \sum_{p=0}^{1} e^{-2\beta \left( \frac{1}{2} + \delta^{N=2}_p \right)} \nl
&=&
2 e^{-\beta} \left[ 1 + \frac{8 \beta {\cal B}_0}{\pi \hbar} \left\{ (1 + \cos \theta ) \left(\beta +2 \log \frac{\hbar }{16} -2 \gamma \right) \mp 2  \pi i \right\} \right] +O({\cal B}_0^2). \nl
\ee
One can immediately find that there only exists the terms proportional to $\cos (Q \theta)$ with $Q\in {\mathbb N}_0$, which means that the contribution from the $Q_p \notin N {\mathbb Z}$ sector is cancelled by other $p$-sectors.

%%%%%%%%%%%%%%%%%%%%%%%%%%%%%%%%%

\subsection{Resurgent structure of the Hilbert space and the partition function}
\label{sec:res-tri}
Finally, we  comment on the resurgent structure of the quantizaion conditions,  the partition function, and fixed topological charge sectors of the partition function (which are the columns of resurgence triangle).

\begin{figure}[t]
    \centering 
    \includegraphics[width=17cm]{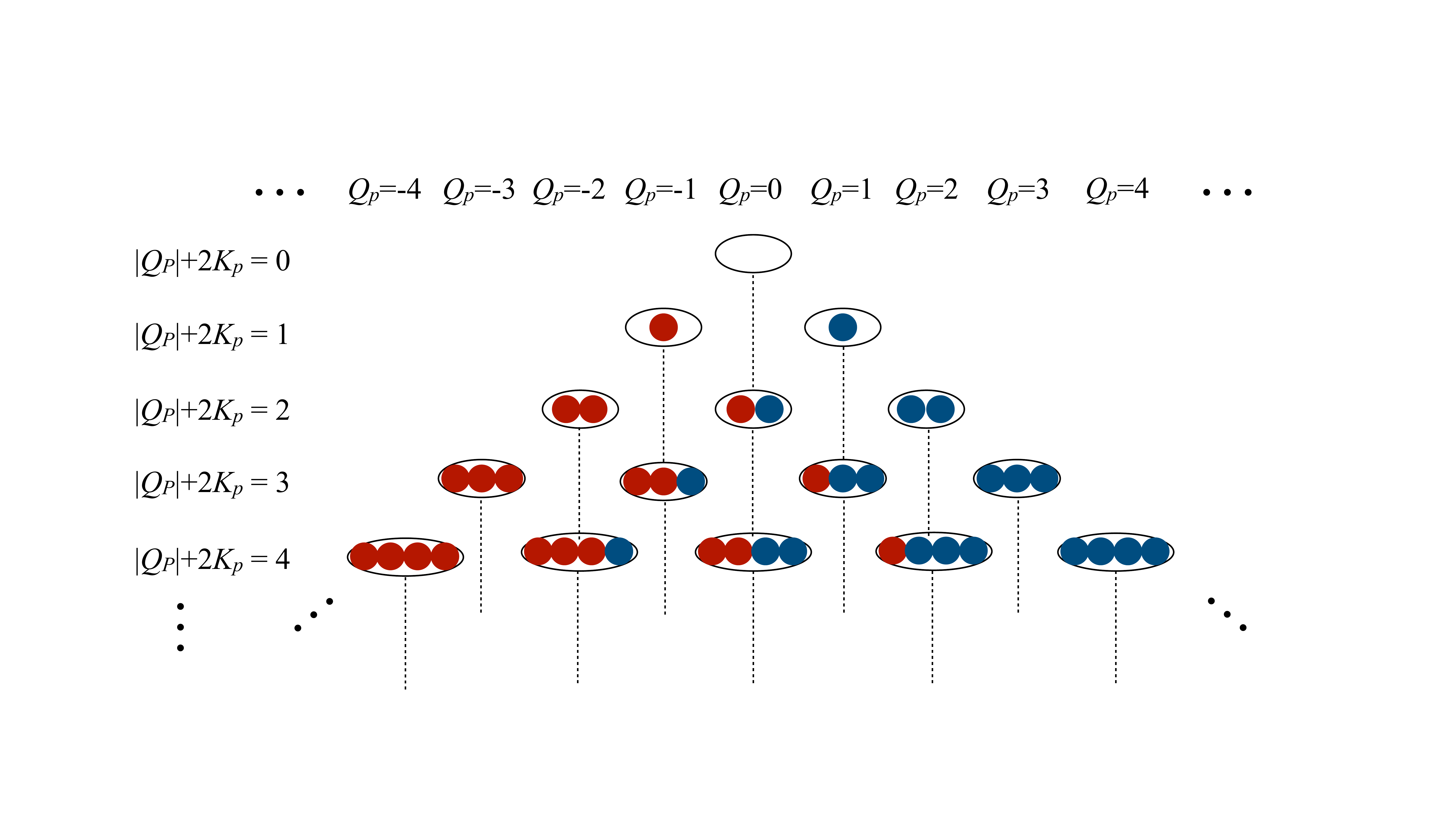}
    \caption{Resurgence triangle exhibiting the complete resurgent structure %in Eq.~(\ref{eq:cres})
    is depicted, where blue points stand for instantons and red points for anti-instantons. We show the structure for each $p$. Thus, there are $N$ copies of this structure in total. By summing up all of the $p$-sectors, the cancellation among the $p$-sectors arises and the contribution only from the $Q_p$-sectors with $Q_p \in N {\mathbb Z}$ remains. }
    \label{fig:res}
\end{figure}

We  first consider the quantization condition and in order to see the implication of resurgence in fixed discrete theta angle $p$,   we rewrite Eq.(\ref{eq:DNN}) as
\be
   {\cal D}^{(N)\pm} &=& \frac{1}{\left( {\cal A}^{\mp 1} {\cal B} \right)^{N/2}} \prod_{p=0}^{N-1} {\cal D}^{(N)\pm}_p \nl
   &=& (\alpha^{\pm})^{N} \prod_{p=0}^{N-1} \left( 1 - \beta^{\pm} e^{+i(\theta + 2 \pi p)/N}\right) \left( 1 - \beta^{\pm} e^{-i(\theta + 2 \pi p)/N}\right), 
\ee
where
\be
&& {\cal D}^{(N)\pm}_{p}:= 1 + {\cal A}^{\mp 1}(1+{\cal B}) - 2 \sqrt{{\cal A}^{\mp 1} {\cal B}} \cos \left( \frac{\theta + 2 \pi p}{N} \right), \label{eq:DNp} \\
&&   \alpha^{\pm} = \xi^{\pm} + \sqrt{(\xi^{\pm})^2 -1}, \qquad \beta^{\pm} = \xi^{\pm} - \sqrt{(\xi^{\pm})^2 -1}, \qquad \xi^{\pm} = \frac{1+{\cal A}^{\pm 1} + {\cal B}}{2\sqrt{{\cal A}^{\pm 1}{\cal B}}}.
\ee
Since the DDP formula gives
\be
 {\cal S}_+[\sqrt{\cal A}]  = {\cal S}_- {\frak S}[\sqrt{\cal A}]   \quad \Rightarrow \quad {\cal S}_+[\xi^{+}] = {\cal S}_-[\xi^{-}] \quad \Rightarrow \quad
 \begin{cases}
   {\cal S}_+[\alpha^{+}] = {\cal S}_-[\alpha^{-}] \\
   {\cal S}_+[\beta^{+}] = {\cal S}_-[\beta^{-}] 
 \end{cases}, 
\ee
where ${\frak S}$ denotes the Stokes automorphism defined as ${\frak S}[\sqrt{\cal A}]=(1 + {\cal B})\sqrt{\cal A}$,
one can easily see that the DDP transformation is closed in  each of $p$-sectors. This is not surprising, as each one of the $p$ sectors corresponds to $N=1$ system with the 
replacement of $\theta \rightarrow \frac{\theta + 2 \pi p}{N}$. 
The energy spectrum obtained by solving the quantization condition directly corresponds to the Hilbert space ${\cal H}_p$, and it means that ${\cal H}_p$ is invariant under the the Stokes automorphism ${\cal S}$.
%and it is irreducible on ${\cal H}_p$.%\textcolor{red}{(also should be labeled by the energy level???)}
At the same time, the partition function can be written through the resolvent method by keeping $(\alpha^{\pm},\beta^{\pm})$ as\footnote{
$\log{\sqrt{{\cal A}^{\mp 1} {\cal B}}}$ in Eq.(\ref{eq:totZ}) is nothing but a convention to make the perturbative part as ${\cal D}_{\cal A}^{\mp}$ in ${\cal D}^{(N)\pm}$.
It does not affect  the resulting partition function because it disappears in the energy integration.
}
\be
 Z^{\pm}(\hbar,\beta) &=& \frac{\beta}{2 \pi i} \int^{\epsilon + i \infty}_{\epsilon - i \infty} \left[ - N \log \left( \sqrt{{\cal A}^{\mp 1}{\cal B}} \alpha^{\pm} \right) + \sum_{p=0}^{N-1} \sum_{Q_p \in {\mathbb Z}\setminus \{0\}} \frac{\left( \beta^{\pm} \right)^{|Q_p|}}{|Q_p|} e^{i(\theta+2 \pi p )Q_p/N} \right] e^{-\beta E} dE \label{eq:totZ} \nl \\ 
&=& \frac{\beta}{2 \pi i} \int^{\epsilon + i \infty}_{\epsilon - i \infty} \left[ - N \log \left( \sqrt{{\cal A}^{\mp 1}{\cal B}} \alpha^{\pm} \right) + \sum_{Q \in {\mathbb Z}\setminus \{0\}} \frac{\left( \beta^{\pm} \right)^{N|Q|}}{|Q|} e^{i Q \theta} \right] e^{-\beta E} dE. \label{eq:totZ2}
\ee
The first and second terms in Eq.(\ref{eq:totZ}) correspond to the $Q_p=0$ and $Q_p \ne 0$ sectors, respectively, and we performed the discrete Fourier transform by summing $p$ up to obtain Eq.(\ref{eq:totZ2}).
Notice that the DDP invariance for the $Q_p=0$ sector can be ensured as
  \be
     {\cal S}_+[Z^{+}_{Q_p=0}(\hbar,\beta)]  &=& -\frac{N\beta}{2 \pi i} \int^{\epsilon + i \infty}_{\epsilon - i \infty} {\cal S}_+ \left[ \log \left( \sqrt{{\cal A}^{-1}{\cal B}} \alpha^{+} \right) \right] e^{-\beta E} dE \nl
%     &=& -\frac{N\beta}{2 \pi i} \int^{\epsilon + i \infty}_{\epsilon - i \infty} {\cal S}_- \left[ \log \left( \frac{\sqrt{{\cal A}^{-1}{\cal B}}}{1+{\cal B}} \alpha^{-} \right) \right] e^{-\beta E} dE \nl
     &=& -\frac{N\beta}{2 \pi i} \int^{\epsilon + i \infty}_{\epsilon - i \infty} {\cal S}_- \left[ \log \frac{1}{{\cal A}(1+{\cal B})} + \log \left( \sqrt{{\cal A}{\cal B}} \alpha^{-} \right)   \right] e^{-\beta E} dE \nl
     &=& -\frac{N\beta}{2 \pi i} \int^{\epsilon + i \infty}_{\epsilon - i \infty} {\cal S}_- \left[ \log \left( \sqrt{{\cal A}{\cal B}} \alpha^{-} \right)   \right] e^{-\beta E} dE \, = \, {\cal S}_-[Z^{-}_{Q_p=0}(\hbar,\beta)].
  \ee
To drive the third line, we have used the fact that $\log \frac{1}{{\cal A}(1+{\cal B})}$ is a holomorphic function of $E$. As emphasized earlier, $p$ has few equivalent interpretation. One is the label of  Bloch momentum   that is eigenstate of the ${\mathbb Z}_N$-translation  symmetry given by $x \rightarrow x+2 p \pi/N$.
The important fact is that the ${\cal D}_p^{(N)\pm}$ gives the $Q_p$-summed partition function which is decomposable into the $Q_p$-sectors for the Stoke automorphism.
It can be seen from the fact that $\sum_{K_p} Z(\beta;\{p,Q_p,K_p\})$ by Eq.(\ref{eq:ZpQK})
is indeed invariant under the DDP transformation and irreducible in the sense that $Z(\beta;\{p,Q_p,K_p\})$ is not invariant.
Therefore, the trans-series structure of the partition function is characterized not only by the topological charge but also the  $\Z_N$ translation symmetry. 
%We can rephrase it as ``the resurgent structure is closed both in the topological sector and in the $N$-shift ($N$-ality) sector.''
%Hence, the group action of the Stokes automorphism ${\cal S}$ on the $(p,Q_p,K_p)$-sector in the partition function can be expressed as
%\be
%&& (p,Q_p,K_p) \xrightarrow[]{\quad {\cal S} \quad} \bigoplus_{K^\prime_p \ge K_p} (p,Q_p,K_p^\prime), \label{eq:cres} 
%\ee
%where $Z_{\rm p}(\hbar,\beta)  \in (p,0,0)$ and  $Z_{\rm np}(\hbar,\beta;\{p,Q_p,K_p\}) \in (p,Q_p,K_p)$ when $|Q_p|+K_p \ne 0$.
This complete resurgent structure of the partition function is exhibited in Fig.~\ref{fig:res}.
We summarize the relationship of the Hilbert space and the partition function below:
\if0
\be
\begin{large}
\begin{tikzcd}
  &  &  {\cal H}_p \arrow[out=120,in=60,loop,"{\frak S}"] \arrow{r}{\bigoplus_p} \arrow[leftrightarrow]{dd}{}{\mbox{ Resolvent}} & {\cal H} \arrow[leftrightarrow]{dd}{}{\mbox{}}  \\
  & & & \\
  {\cal Z}_{p,Q_p,K_p} \arrow[bend right]{r}[swap]{\frak S} \arrow{r}{}{\bigoplus_{K_p}} &{\cal Z}_{p,Q_p} \arrow[out=120,in=60,loop,"{\frak S}"] %\arrow[out=120,in=60,loop,"{\cal S}"]
  \arrow{r}{\bigoplus_{Q_p}}  \arrow{dr}[swap]{\sum_p}&  {\cal Z}_p \arrow{r}{\bigoplus_p} & {\cal Z} \\
  && {\cal Z}_{0,QN} \arrow{ur}[swap]{\sum_Q}&
\end{tikzcd}
\end{large}
\ee
\fi
\be
\begin{large}
\begin{tikzcd}[row sep=large, column sep=large]
  &  &  {\cal H}_p \arrow[out=120,in=60,loop,"{\frak S}"] \arrow{r}{\bigoplus_p} \arrow[leftrightarrow]{d}{}{\mbox{ Resolvent}} & {\cal H} \arrow[leftrightarrow]{d}{}{\mbox{}} \\ [4.5ex]
  %& & & \\
  {\cal Z}_{p,Q_p,K_p} \arrow[bend right]{r}[swap]{\frak S} \arrow{r}{}{\bigoplus_{K_p}} &{\cal Z}_{p,Q_p} \arrow[out=120,in=60,loop,"{\frak S}"] %\arrow[out=120,in=60,loop,"{\cal S}"]
  \arrow{r}{\bigoplus_{Q_p}}  \arrow{d}[swap]{\sum_p}&  {\cal Z}_p \arrow{r}{\bigoplus_p} & {\cal Z} \\ [4ex]
  &{\cal Z}_{0,QN} \arrow{urr}[swap]{\sum_Q} & &
\end{tikzcd}
\end{large} \label{eq:comm_graph}
\ee
where $Z^\pm(\hbar,\beta;\{p,Q_p,K_p \}) \in {\cal Z}_{p,Q_p,K_p}$, $Z^\pm(\hbar,\beta) \in {\cal Z}$, and
\small
\be
&& {\cal H}_p := \{\,  | \psi \rangle \in L^2({\mathbb R})
\, : \, \widehat{H} | \psi \rangle = E_{n,p} | \psi \rangle \ \mbox{where $E_{n \in {\mathbb N_0},p}$ are solutions of ${\cal D}^{N(\pm)}_{p}(E)=0$} \, \}, \nl
%&& {\cal H} := \bigoplus_{p=0}^{N-1} {\cal H}_p, \\
&&  {\cal Z}_{p,Q_p,K_p} := \{ {\mathbb C}[[\hbar,\frac{e^{-\frac{S_B}{2\hbar}}}{\hbar^{1/2}}, \log \hbar]] \,:\, e^{i(Q_p + 2 \pi p)/N} \int^{\epsilon + i \infty}_{\epsilon - i \infty} dE e^{-\beta E} \, {\cal B}^{|Q_p|/2+K_p} {\mathbb C}[[{\cal A}^{\mp 1}]] \}. 
%&& {\cal Z}_{p,Q_p} := \bigoplus_{K_p=0}^\infty {\cal Z}_{p,Q_p,K_p}, \qquad  {\cal Z}_{p} := \bigoplus_{Q_p=0}^\infty {\cal Z}_{p,Q_p}, \qquad  {\cal Z} := \bigoplus_{p=0}^{N-1} {\cal Z}_{p}.
\ee
\normalsize
From Eq.(\ref{eq:comm_graph}) one can immediately claim that the resurgence in Eq.(\ref{eq:tpf2}) is closed within each fixed topological sector of $\tilde{Z}_p$ (and $Z_\ell$ via the Fourier transform). \\

\noindent
{\bf Alternative derivation:}   Consider $T_1$ model with theta angle.  Exact quantization condition for this system is given in 
 \eqref{eq:D_S1}. The solution of the  $D^\pm(E, \theta)=0$  is in correspondence with the Hilbert space ${\cal H}_{\theta}$, based on the $\theta$ vacuum, 
   \begin{align}
|  \Psi_\theta \rangle  =  \sum_{n \in \Z} e^{i n \theta}   | 0_n \rangle
\end{align}  
 where  $| 0_n \rangle$ is the harmonic ground state and its copies under large gauge transformation.    This perspective follows from the fact that 
 to obtain  $T_1$ model,   we  gauged $\Z$ translation symmetry for  the particle  on a line  $x \in \mathbb R$, and obtained
  $x \in S^1 =  \mathbb R/  2 \pi  \mathbb Z $ with $N=1$. (For detailed discussion of particle on a line in periodic potential vs. particle on a circle, see \cite{Rajaraman:1982is})

 We showed that the condition  $D^\pm(E, \theta)=0$ is invariant under  left/right Borel resummation \eqref{invar}. This implies that the partition function of the $T_1$ system 
% A good way to think about this system is to start with particle on 
%   correspond to eigenenergies of the system in 
  \begin{align}
Z(\theta) =  \sum_{Q \in \Z}        e^{ i  \theta Q} \int_{x(\beta) = x(0) + 2 \pi Q} Dx \; e^{-S[x]}  \;\;  \equiv \;\;    \sum_{Q \in \Z}        e^{ i  \theta Q}  Z_{Q}     
\end{align}  
is invariant under left/right Borel  resummation.     The Fourier coefficients correspond to the  twisted partition functions,  
  \begin{align}
Z_{Q}   = \tr[(\mathsf U)^Q e^{-\beta H}]  = \frac{1}{2\pi} \int_{0}^{2 \pi} d\theta  \;      e^{- i  \theta Q} Z(\theta) 
  \end{align} 
  which are in one-to-one correspondence with the columns of resurgence triangle. The linearity of the Fourier transform implies that each column of the resurgence triangle, i.e, each fixed topological charge sector,  is closed under resurgence.  Ambiguities of Borel resummation of perturbation theory around the  multi-instanton  $[I^k] \sim e^{-k S_I/\hbar } P(\hbar) $  are cured 
  by the ambiguity in the amplitude of $[I^{k+1} \bar I]_{\pm}, [I^{k+2}  \bar I^{2}  ]_{\pm}, \ldots $ events etc.

%%%%%%%%%%%%%%%%%%%%%%%%%%%%%%%%%

\section{Summary and discussion}
\label{sec:SD}
We have investigated quantum mechanical systems of a particle on  $S^{1}$ in the presence of a  periodic potential  with $N$-minima $(N=1,2 \ldots) $
 by the exact-WKB method. 
We used   Stokes graphs with  both Airy  and  degenerate Weber type building blocks, and determined exact quantization conditions. 
By using the DDP formula that related the  perturbative and non-perturbative cycles, we showed invariance of the quantization condition under Stokes automorphism. 
This implies that all orders perturbative/non-perturbative resurgent cancellations is implicit in the quantization condition and partition function.

The implication of our result for the Gutzwiller trace formula are also discussed.  In particular, our construction 
 identifies prime periodic orbits that enter to the trace formula, leading to an understanding of its $\theta$ dependence.  
  Exact-WKB analysis correctly produces the conjectured quantization condition for $N=1$. 
The symbolic forms of the Fredholm determinant obtained by Airy-type~(\ref{eq:D_S1}) and by degenerate Weber-type~(\ref{eq:cal_B}) coincide. 
%The resurgent structure is obtained by using the DDP formula in both cases.
Our result obtained by the degenerate Weber-type  correctly reproduces the energy eigenvalues conjectured by Zinn-Justin \cite{ZinnJustin:2004ib}, and obtained earlier by using uniform WKB method in \cite{Dunne:2014bca} for $N=1$. For general $N$, the resurgent structure is closed in the eigenspaces of $\Z_N$-shift symmetry,  i.e, $\widetilde Z_p$ associated with ${\cal H}_p$   subspace of the Hilbert space. The Fourier transform of this relation gives us the result that resurgent structure is closed  
in fixed topological charge sectors. This implies that the ambiguity of perturbation theory around an instanton $[I]$ is cured  by $[I I \bar I]_{\pm},  \; [I  I I \bar I \bar I]_{\pm}$  etc.   which lives on the same topological charge sector. According to DDP formula, this structure  is true on all columns of resurgence triangle. 
%We have shown that the resurgent structure is closed both in the $N$-shift sector \textit{and} in the topological sector.

Furthermore, we have also shown that exact quantization condition naturally captures the mixed 't Hooft anomaly  \cite{Gaiotto:2017yup}  or global inconsistency between $\Z_N$ translation  symmetry and charge conjugation symmetry $C$ \cite{Kikuchi:2017pcp}. 

%To sum up, we have obtained the quantization condition and have shown the detailed resurgent structure in quantum mechanics of a particle on $S^{1}$, where we have used the dictionary connecting the Airy-type and the degenerate-Weber-type Stoke graphs. We also exhibit that the exact-WKB analysis is a powerful tool to study physical problems in the systems. 
Below, we  list few  topics to which we may be able to apply our methodology:

\begin{itemize}

	\item
	Tilted periodic-potential quantum mechanics, including supersymmetric and quasi-exactly solvable  cases.

	\item 
	Constructive resurgence between perturbative/non-perturbative sectors \cite{Dunne:2013ada, Alvarez3} in $S^1$ quantum mechanics. This type of resurgence connects 
	different topological sectors to each other, unlike the one we discussed here, which always take place within a fixed topological sector.

	\item Compactification of QFTs with background fluxes  down to quantum mechanics, such as the ones discussed in \cite{Unsal:2020yeh}. 
%	Low-dimensional quantum field theories, where we may consider $S^1$ compactification to make the theories reduced to quantum mechanics. 
	
	\item 
	Schwinger mechanism for time-dependent electric field \cite{Taya:2020dco}, where the Klein-Gordon equation in scalar QED has a Schr\"{o}dinger-equation form.
	
\end{itemize}

%Apart from these possibilities,we consider that there could be lots of physical topics we can apply the exact-WKB for quantum mechanics with periodic potential.

Most of readers may be interested in the application of exact-WKB to quantum field theory (not to the reduced quantum mechanics).
Toward this goal, we first need to investigate the exact-WKB analysis for quantum mechanics with multiple degrees of freedom, e.g. 
After completing such  extension, we will consider taking the limit of infinite degrees of freedom, which could have  an implication to  quantum field theory.
%This project will be embarked on in our future works.

%%%%%%%%%%   ACKNOWLEDGMENTS   %%%%%%%%%%
\begin{acknowledgements}
    The authors especially thank O. Morikawa and students of E-lab for fruitful discussion on quantization conditions for $S^{1}$ quantum mechanics.
	T.\ M. is supported by the Japan Society for the Promotion of Science (JSPS) Grant-in-Aid for Scientific Research (KAKENHI) Grant Numbers 18H01217 and 19K03817.
	S.~K. is supported by the Polish National Science Centre grant 2018/29/B/ST2/02457. 
	The original questions related to the present work were posed in ``RIMS-iTHEMS International Workshop on Resurgence Theory" 
    at RIKEN, Kobe in 2017. The authors are grateful to the organizers and participants of the workshop.
	M.~U. acknowledges support from U.S. Department of Energy, Office of Science, Office of Nuclear Physics under Award Number DE-FG02-03ER41260.
	The authors thank Yukawa Institute for Theoretical Physics at Kyoto University. Discussions during the YITP-RIKEN iTHEMS workshop YITP-T-20-03 on "Potential Toolkit to Attack Nonperturbative Aspects of QFT -Resurgence and related topics-" were useful to complete this work.
\end{acknowledgements}

%%%%%%%%%%%%%%%%%%%%%%

\appendix 

%%%%%%%%%%%%%%%%%%%%%%%%%%%%%%%%%%%%%
\section{The degenerate Weber equation} \label{sec:dweq}
%%%%%%%%%%%%%%%%%%%%%%%%%%%%%%%%%%%%%

In   appendix~\ref{sec:der_DWconn},   we  review the derivation of the connection formula for the degenerate-Weber type Stokes graph. 
See pages 28-40 of Ref.~\cite{Kawai1} for the proof of this construction for the Airy type building blocks of Stokes graph (corresponding 
to $\widehat{Q}(y(x,\hbar),\hbar)= y(x, \hbar)$ below). Appendix~\ref{sec:der_DWconn}  is the generalization of that local/global relations by using  degenerate 
Weber type building blocks of Stokes graph. 
Appendix~\ref{sec:dic_A_DW} is devoted to  the construction of dictionary between  the Airy-type and the degenerate Weber-type building blocks.   
In order to avoid confusion, we summarized the notation for global and local variables in Tab.\ref{tab:notation_glo_loc}, which are used below.

%%%%%%%%%%%%%%%%%%%%%%%%%%%%%%%%%%%%%
\subsection{Derivation of the connection formula} \label{sec:der_DWconn}
%%%%%%%%%%%%%%%%%%%%%%%%%%%%%%%%%%%%%

As we discussed through the paper,  given a classical potential, we can immediately obtain the Stokes graph. The building blocks of the Stokes graph are either Airy type  or degenerate Weber type building  blocks.  In this appendix, we  review the derivation of connection formula of the degenerate Weber(DW) equation.   
In the construction of  exact quantization conditions, three important matrices play a role. Connection matrices used for passage through Stokes line, normalization matrix (or Voros multiplier) which accounts for the change of turning point, and  branchcut matrices which account for the passage through a branchcut. The combination of these and 
a global boundary condition on WKB wave function is the exact quantization condition.

%%%%%%%%%%%%%%%%%%%%%%%%%%%%%%%%%%%%%%%%%%%%%%%%%%
\setlength{\tabcolsep}{10pt} 
\renewcommand{\arraystretch}{1.5}
\begin{table}[t]
\centering
 \begin{tabular}{|c|c|c|c|c|} 
 \hline
  & Wavefunction & Classical potential  & Riccati var. & Energy\\  \hline
 Global & $\psi(x,\hbar)$ & $Q(x,\hbar) = 2(V(x)-E)$ & $S(x,\hbar)$ & $E$ \\
 Local & $\widehat{\psi}(y,\hbar)$ & $\widehat{Q}(y,\hbar)= \frac{y^2}{4} -  \hbar \kappa $ & $\widehat{S}(y,\hbar)$ & $\kappa/2$ \\ 
 \hline
 \end{tabular}
 \caption{Notation table for global and local variables}
 \label{tab:notation_glo_loc}
\end{table}
%%%%%%%%%%%%%%%%%%%%%%%%%%%%%%%%%%%%%%%%%%%%%%%%%%

In order to obtain the formula, one has to take the two steps: Firstly obtaining the connection formula for the local coordinate, and then lifting them up to the \textit{global} coordinate.
The following relations which are equivalent to each others are important:
\be
&(1)&   \  {Q}({x},\hbar)=  \left(\frac{\pd y({x},\hbar)}{\pd {x}}\right)^2  \widehat{Q}(y(x,\hbar),\hbar) - \frac{\hbar^{2}}{2} \{ y(x,\hbar);{x} \}, \label{eq:QQ_dW} \\
&(2)& \ S^{(\pm)}(x,\hbar) = \frac{\pd y(x,\hbar)}{\pd x} \cdot \widehat{S}^{(\pm)}(y(x,\hbar),\hbar) - \frac{1}{2}  \frac{\frac{\pd^2 y({x},\hbar)}{\pd {x}^2}}{\frac{\pd y({x},\hbar)}{\pd {x}}}, \\
&(3)& \ S_{\rm odd}(x,\hbar) = \frac{\pd y(x,\hbar)}{\pd x} \cdot \widehat{S}_{\rm odd}(y(x,\hbar),\hbar), \label{eq:Sod_Sod}
\ee
where
\be
&& \{ y;{x} \} = \frac{\frac{\pd^3 y({x},\hbar)}{\pd {x}^3}}{\frac{\pd y({x},\hbar)}{\pd {x}}} - \frac{3}{2} \left( \frac{\frac{\pd^2 y({x},\hbar)}{\pd {x}^2}}{\frac{\pd y({x},\hbar)}{\pd {x}}} \right)^2, \\
&& y(x,\hbar) = \sum_{n=0}^{+\infty}  y_{n}(x) \hbar^n,
\ee
$x$ and $y$ are the global and local coordinates, respectively, and $S(x,\hbar)$/$\widehat{S}(y,\hbar)$ is the global/local asymptotic solution of the Riccati equation with a global/local potential $Q(x,\hbar)$/$\widehat{Q}(y,\hbar)$, 
\be
&& S(x,\hbar) = \sum_{n=-1}^{\infty} S_n(x) \hbar^n, \\
&& S^{(\pm)}_{-1}(x) = \pm \sqrt{Q_0(x)}, \\
&& 2 S^{(\pm)}_{-1}(x) S^{(\pm)}_{n}(x) + \sum_{k,\ell = 0}^{k+\ell=n-1} S^{(\pm)}_{k}(x) S^{(\pm)}_{\ell}(x) + \frac{d S^{(\pm)}_{n-1}(x)}{dx} = Q_{n+1}(x), \quad n \in {\mathbb N}_0, \\
&& S_{\rm odd}(x,\hbar)= \frac{1}{2} \left( S^{(+)}(x,\hbar) - S^{(-)}(x,\hbar) \right), \qquad S_{\rm even}(x,\hbar)= \frac{1}{2} \left( S^{(+)}(x,\hbar) + S^{(-)}(x,\hbar) \right). \nl 
\ee
From the above relationships, the global and local wavefunctions, $\psi(x,\hbar)$ and $\widehat{\psi}(y,\hbar)$, can be connected as
\be
\psi_{\pm}(x,\hbar) = C_{\pm}(\hbar) \left(  \frac{\pd y(x,\hbar)}{\pd x} \right)^{-1/2}  \widehat{\psi}_{\pm}(y(x,\hbar),\hbar), \label{eq:psi_tilpsi}
\ee
where
\be
C_{\pm}(\hbar) = \sum_{n=0}^{+\infty}  C_{\pm, n} \hbar^n. \label{eq:Cpm_eta}
\ee

Let us start with the degenerate Weber  equation given by
\be
&&\left[ - \hbar^{2} \frac{\pd^2}{\pd y^2} + \widehat{Q}(y,\hbar) \right] \widehat{\psi}(y,\hbar) = 0, \qquad  
\widehat{Q}(y,\hbar) = \frac{y^2}{4} -  \hbar \kappa, \qquad \kappa \in {\mathbb R},\label{eq:DegWeb_psi}
\ee
where $y \in {\mathbb C}$ is the local coordinate and $\widehat{\psi}$ is a local wavefunction.
By solving the Riccati equation recursively, the formal solution is defined as
\be
\widehat{\psi}_{\pm}(y,\hbar) %&=& \frac{e^{ \pm \int^{y}_{0}dy \, \left( \hbar^{-1} \widehat{S}_{\rm odd,-1}(y) + \widehat{S}_{\rm odd,0}(y) \right) \pm \int^{y}_{\infty} dy \, \left( \widehat{S}_{\rm odd}(y,\hbar) - \hbar^{-1} \widehat{S}_{\rm odd,-1} (y) - \widehat{S}_{\rm odd,0}(y) \right) }}{\sqrt{\widehat{S}_{\rm odd}(y,\hbar)}} \nl
&=&  \frac{y^{\mp \kappa} e^{\pm \hbar^{-1} y^2/4}}{\sqrt{\widehat{S}_{\rm odd}(y,\hbar)}} \exp \left [ \pm \int^{y}_{\infty} dy \, \left( \widehat{S}_{\rm odd}(y,\hbar) - \hbar^{-1} \widehat{S}_{\rm odd,-1} (y) - \widehat{S}_{\rm odd,0}(y) \right) \right],  \label{eq:formal_psi_hat}
\ee
where
\be
\widehat{S}_{\rm odd,-1}(y) = \frac{y}{2}, \qquad \widehat{S}_{\rm odd,0}(y) = -\frac{\kappa}{y},
\ee
and other $\widehat{S}_{{\rm odd},n>0}(y)$ can be computed in the similar way. 
In Eq.(\ref{eq:formal_psi_hat}), we took a reference point of the normalization at $y = \infty$ for $\widehat{S}_{{\rm odd},n>0}(y)$.
For convenience, we redefine the wave function as
\be
\widehat{u}_{\pm}(y,\hbar) := \hbar^{\pm \kappa/2} \widehat{\psi}_{\pm}(y,\hbar). \label{eq:u_eta}
\ee
The asymptotic solution can be obtained as \footnote{At this stage, it is useful to realize that the construction based on degenerate Weber type building blocks is 
intimately related to uniform-WKB approach.  In uniform WKB, one starts with an ansatz $\psi(y) = \frac{1}{ \sqrt { u'(y)}} D_\nu (\frac{1}{\hbar}  u(y)) $  
where $D_\nu$ is parabolic cylinder (Weber) function  and $\nu$ is ansatz parameter. Then,  one expands  
 $u(y)$ to a formal power series $u(y)= u_0(y) + \hbar u_1 (y) + \hbar^2 u_2 (y) + \ldots$, see e.g. \cite{Dunne:2014bca}. In certain sense, these two approach are very similar, both take advantage of the fact that in the $\hbar \rightarrow 0$ limit,   the system would be described by harmonic minima. This is the point that actually generates 
 the differences compared to Airy-type decomposition of Stokes graph, and makes it more suitable to obtain spectral information from the Weber type decomposition. }

\be
\widehat{u}_{\pm}(y,\hbar) &=& \sqrt{2 y}  \; e^{\pm \hbar^{-1} y^2/4} \sum_{n=0}^{+\infty} \frac{\widehat{u}^{\pm}_{n}}{y^{1 \pm \kappa + 2n}} \hbar^{(1\pm \kappa)/2 + n}. \label{eq:upm_asym}
\ee
Substituting Eq.(\ref{eq:upm_asym}) into Eq.(\ref{eq:DegWeb_psi}) gives the recursion relation for the coefficients $\widehat{u}^{\pm}_n$ as
%\be
%({\rm ODE}) &=& 
%&&\left[ \mp 2 n \hbar^{-1}    + \frac{ \left(1/2 \pm \kappa + 2 n \right)\left(3/2 \pm \kappa + 2 n \right)}{y^2}  \right] \nl
%&& \cdot \sqrt{2 y} e^{\pm \hbar^{-1} y^2/4} \sum_{n=0}^{+\infty} \frac{u^{\pm}_{n}}{y^{1 \pm \kappa + 2n}} \hbar^{(1\pm \kappa)/2 + n} \ = \ 0,
%\ee
\be
 2 (n+1) \widehat{u}^{\pm}_{n+1}  = \pm \left( \frac{1}{2} \pm \kappa + 2 n \right) \left( \frac{3}{2} \pm \kappa + 2 n \right) \widehat{u}^{\pm}_{n} \qquad \mbox{with} \quad \widehat{u}^{\pm}_{0}= 1,
\ee
and the coefficients are obtained as
\be
\widehat{u}^{\pm}_{n} = (\pm 2)^{n-2} \frac{(1/2 \pm \kappa)(3/2 \pm \kappa) (5/4 \pm \kappa/2)_{n-1} (7/4 \pm \kappa/2)_{n-1}}{\Gamma(n+1)},
\ee
where $(x)_{n}=\Gamma(x+n)/\Gamma(x)$ is the Pochhammer symbol.
%It is notable that
%\be
%u^{\pm}_n  = (-1)^{n} \left. u^{\mp}_{n} \right|_{\kappa \rightarrow -\kappa}. \label{eq:u_refl}
%\ee
Acting the Borel transform ${\cal B}$ to the asymptotic solution expanded by $\hbar$ gives
\be
\widehat{u}_{B\pm}(y,\xi) &:=& {\cal B}[\widehat{u}_{\pm}](y,\xi) \nl
&=& y^{-3/2} 2^{1 \mp \kappa/2}  \sum_{n=0}^{+\infty} \frac{\widehat{u}^{\pm}_{n}}{2^n \Gamma \left( \frac{1 \pm \kappa}{2} + n \right)} \left( \frac{2\xi}{y^2} \pm \frac{1}{2}\right)^{(-1 \pm \kappa)/2 + n}.
\ee
By taking the summation, it can be expressed as
\be
&& \widehat{u}_{B \pm}(y,\xi) = A_{\pm} y^{-3/2} (\pm s_{\pm})^{-\gamma_{\mp}} F(\alpha_{\pm},\beta_{\pm};\gamma_{\pm};s_{\pm}), 
\ee
where $F(\alpha,\beta;\gamma;s)$ is the Gauss hypergeometric function and
\be
&& s_{\pm} =
\begin{cases}
  s & \mbox{for } s_+ \\
  1-s & \mbox{for } s_-
\end{cases} \quad \mbox{ with } \  s=\frac{2 \xi}{y^2} + \frac{1}{2}, \\
&& A_{\pm} = \frac{2^{1/2+\gamma_{\mp}}}{\Gamma(\gamma_{\pm})}, \quad  \alpha_\pm = \frac{1}{4} \pm \frac{\kappa}{2}, \quad \beta_\pm = \frac{3}{4} \pm \frac{\kappa}{2}, \quad \gamma_\pm = \frac{1}{2} \pm \frac{\kappa}{2}.
\ee
From identities of the hypergeometric function, one finds that
\be
F(\alpha_{\pm},\beta_{\pm};\gamma_{\pm};s_{\pm}) &=&  s_{\mp}^{-\gamma_{\pm}} \frac{\Gamma(\gamma_{\pm})^2}{\Gamma(\alpha_{\pm}) \Gamma(\beta_{\pm})} F(1/4,-1/4;\gamma_{\mp};s_{\mp}) \nl
&& +  \frac{\Gamma(\gamma_{\pm}) \Gamma(-\gamma_{\pm})}{\Gamma(1/4) \Gamma(-1/4)} F(\alpha_{\pm},\beta_{\pm};1+\gamma_{\pm};s_{\mp}) \nl
&=&  s_{\mp}^{-\gamma_{\pm}} \frac{\Gamma(\gamma_{\pm})^2}{2^{1-2 \alpha_{\pm}} \sqrt{\pi} \Gamma(2 \alpha_\pm)} s_{\pm}^{\gamma_\mp} F(\alpha_{\mp},\beta_{\mp};\gamma_{\mp};s_{\mp}) \nl
&& +  \frac{\Gamma(\gamma_{\pm}) \Gamma(-\gamma_{\pm})}{\Gamma(1/4) \Gamma(-1/4)} F(\alpha_{\pm},\beta_{\pm};1+\gamma_{\pm};s_{\mp}) .
\ee
The second term is irrelevant because it has no singularities.
Thus, by picking up only the first term, one obtains
\be
&& A_{\pm} y^{-3/2} (\pm s_{\pm})^{-\gamma_{\mp}} s_{\mp}^{-\gamma_{\pm}} \frac{\Gamma(\gamma_{\pm})^2}{2^{1-2 \alpha_{\pm}} \sqrt{\pi} \Gamma(2 \alpha_\pm)} s_{\pm}^{\gamma_\mp} F(\alpha_{\mp},\beta_{\mp};\gamma_{\mp};s_{\mp}) \nl
&=& A_{\mp} y^{-3/2} (\mp s_{\mp})^{-\gamma_{\pm}} \frac{\sqrt{\pi}(-1)^{-(1+\kappa)/2}}{\sqrt{2} \cos (\pi \kappa/2) \Gamma(2\alpha_{\pm})}  F(\alpha_{\mp},\beta_{\mp};\gamma_{\mp};s_{\mp})\nl
&=& \frac{\sqrt{\pi}(-1)^{-(1+\kappa)/2}}{\sqrt{2} \cos(\pi \kappa/2) \Gamma(2\alpha_{\pm})}  \widehat{u}_{B\mp}(y,\xi).
\ee

By acting the Laplace integration and taking the Hankel contour, one obtains
\be
&& \int_{C} d\xi \,  e^{- \xi \eta} \widehat{u}_{B\pm}(y,\xi) \nl
&=&   \int_{\pm x^2/4}^{+\infty} d\xi \, e^{- \xi \eta} \frac{\sqrt{\pi} e^{\pi i(1+ \kappa)/2}}{\sqrt{2} \cos(\pi \kappa/2) \Gamma(1/2 \pm \kappa)} \widehat{u}_{B\mp}(y,\xi) (1-e^{-\pi i (1 \pm \kappa)}) \nl
&=&   \int_{\pm y^2/4}^{+\infty} d\xi \, e^{- \xi \eta} \frac{i \sqrt{2 \pi} e^{(1\mp 1)\pi i \kappa/2} }{\Gamma(1/2 \pm \kappa)} \widehat{u}_{B\mp}(y,\xi) \qquad \mbox{ for } \ {\rm IV \rightarrow I}.
\ee

In order to obtain the connection formula passing other Stokes curve, it is convenient to consider analytic continuation for $y$ by
\be
\widehat{u}_{\pm}(y,\hbar) &\xrightarrow{y \rightarrow e^{+ \pi i/2} y}&  e^{- \pi i (1/2 \pm \kappa)/2} \widehat{u}_{\mp}(y,\hbar), \\
\widehat{u}_{\pm}(y,\hbar) &\xrightarrow{y \rightarrow e^{- \pi i} y}&  e^{+ \pi i (1/2 \pm \kappa)} \widehat{u}_{\pm}(y,\hbar), \\
\widehat{u}_{\pm}(y,\hbar) &\xrightarrow{y \rightarrow e^{-\pi i/2} y}&  e^{+ \pi i (1/2 \pm \kappa)/2} \widehat{u}_{\mp}(y,\hbar).
\ee
By repeating the similar above procedure and taking into account Eq.(\ref{eq:u_eta}), the connection formula for Fig.\ref{fig:deg_web_loc} is given by
\be  \label{eq:connect}
\qquad \ {\rm IV \rightarrow I} \ : \ && \Delta_{z=\pm y^2/4} \widehat{\psi}^{\rm IV}_{B\pm}(y,\xi) =  i \frac{\sqrt{2 \pi} e^{+(1 \mp 1)\pi i \kappa/2}}{\Gamma(1/2 \pm \kappa )} \hbar^{\mp \kappa} \widehat{\psi}^{\rm I}_{B\mp}(y,\xi), \label{eq:dw_con_41_p} \\ 
\qquad \ {\rm I \rightarrow II} \ : \ && \Delta_{z=\mp y^2/4} \widehat{\psi}^{\rm I}_{B\mp}(y,\xi) =  i \frac{ \sqrt{2 \pi} e^{+(1 \pm 1)\pi i \kappa/2}}{\Gamma(1/2 \mp \kappa )} \hbar^{\pm \kappa} \widehat{\psi}^{\rm II}_{B\pm}(y,\xi), \\  
\qquad \ {\rm II \rightarrow III} \ : \ && \Delta_{z=\pm y^2/4} \widehat{\psi}^{\rm II}_{B\pm}(y,\xi) =  i \frac{ \sqrt{2 \pi} e^{  -(3 \pm 1) \pi i \kappa/2 }}{\Gamma(1/2 \pm \kappa )} \hbar^{\mp \kappa} \widehat{\psi}^{\rm III}_{B\mp}(y,\xi), \\  
\qquad \ {\rm III \rightarrow IV} \ : \ && \Delta_{z=\mp y^2/4} \widehat{\psi}^{\rm III}_{B-}(y,\xi) =  i \frac{ \sqrt{2 \pi} e^{-(3 \mp 1) \pi i \kappa/2}}{\Gamma(1/2 \mp \kappa )} \hbar^{\pm \kappa} \widehat{\psi}^{\rm IV}_{B \pm}(y,\xi),  
\ee
where $\Delta_{z}$ is the Alien derivative at $\xi = z$.

\eqref{eq:connect} can be viewed as the connection formula where connection formula for the  
Now, we are ready to obtain the global connection formula for a generic potential by  using Eq.(\ref{eq:psi_tilpsi}). The result is 
\be
 &\mbox{ For }& \ {\rm Fig.\ref{fig:deg_web_loc} \, (Left)} : \  \nl
&& \begin{pmatrix}
     \psi^{\rm IV}_{+}(x,\hbar) \\
     \psi^{\rm IV}_{-}(x,\hbar)
   \end{pmatrix} =
\begin{pmatrix}
  1  & & i \frac{C_{+}^{{\rm IV} \rightarrow {\rm I}}(\hbar)}{C_{-}^{{\rm IV} \rightarrow {\rm I}}(\hbar)} \frac{\sqrt{2 \pi} \hbar^{-F(\hbar)}}{\Gamma(1/2 + F(\hbar) )}   \\
0  & & 1
\end{pmatrix}
\begin{pmatrix}
     \psi^{\rm I}_{+}(x,\hbar) \\
     \psi^{\rm I}_{-}(x,\hbar)
\end{pmatrix}, \\
&& \begin{pmatrix}
     \psi^{\rm I}_{+}(x,\hbar) \\
     \psi^{\rm I}_{-}(x,\hbar)
   \end{pmatrix} =
\begin{pmatrix}
 1  & & 0 \\
 i \frac{C_{-}^{{\rm I} \rightarrow {\rm II}}(\hbar)}{C_{+}^{{\rm I} \rightarrow {\rm II}}(\hbar)} \frac{\sqrt{2 \pi} e^{+ \pi i F(\hbar)} \hbar^{+F(\hbar)}}{\Gamma(1/2 - F(\hbar) )} & & 1
\end{pmatrix}
\begin{pmatrix}
     \psi^{\rm II}_{+}(x,\hbar) \\
     \psi^{\rm II}_{-}(x,\hbar)
\end{pmatrix}, \\
&& \begin{pmatrix}
     \psi^{\rm II}_{+}(x,\hbar) \\
     \psi^{\rm II}_{-}(x,\hbar)
   \end{pmatrix} =
\begin{pmatrix}
  1  & & i \frac{C_{+}^{{\rm II} \rightarrow {\rm III}}(\hbar)}{C_{-}^{{\rm II} \rightarrow {\rm III}}(\hbar)} \frac{\sqrt{2 \pi} e^{-2 \pi i F(\hbar)} \hbar^{-F(\hbar)}}{\Gamma(1/2 + F(\hbar) )}   \\
0  & & 1
\end{pmatrix}
\begin{pmatrix}
     \psi^{\rm III}_{+}(x,\hbar) \\
     \psi^{\rm III}_{-}(x,\hbar)
\end{pmatrix}, \\
&& \begin{pmatrix}
     \psi^{\rm III}_{+}(x,\hbar) \\
     \psi^{\rm III}_{-}(x,\hbar)
   \end{pmatrix} =
\begin{pmatrix}
 1  & & 0 \\
 i \frac{C_{-}^{{\rm III} \rightarrow {\rm IV}}(\hbar)}{C_{+}^{{\rm III} \rightarrow {\rm IV}}(\hbar)} \frac{\sqrt{2 \pi} e^{- \pi i F(\hbar)}\hbar^{+F(\hbar)}}{\Gamma(1/2 - F(\hbar) )} & & 1
\end{pmatrix}
\begin{pmatrix}
     \psi^{\rm IV}_{+}(x,\hbar) \\
     \psi^{\rm IV}_{-}(x,\hbar)
\end{pmatrix}, 
\ee
%%%%%%%%%%%%%
\be
 &\mbox{ For }& \ {\rm Fig.\ref{fig:deg_web_loc} \, (Right)} : \  \nl
&& \begin{pmatrix}
     \psi^{\rm IV}_{+}(x,\hbar) \\
     \psi^{\rm IV}_{-}(x,\hbar)
   \end{pmatrix} =
\begin{pmatrix}
 1  & & 0 \\
 i \frac{C_{-}^{{\rm IV} \rightarrow {\rm I}}(\hbar)}{C_{+}^{{\rm IV} \rightarrow {\rm I}}(\hbar)} \frac{\sqrt{2 \pi} e^{+ \pi i F(\hbar)} \hbar^{+F(\hbar)}}{\Gamma(1/2 - F(\hbar) )} & & 1
\end{pmatrix}
\begin{pmatrix}
     \psi^{\rm I}_{+}(x,\hbar) \\
     \psi^{\rm I}_{-}(x,\hbar)
\end{pmatrix}, \\
&& \begin{pmatrix}
     \psi^{\rm I}_{+}(x,\hbar) \\
     \psi^{\rm I}_{-}(x,\hbar)
   \end{pmatrix} =
\begin{pmatrix}
  1  & & i \frac{C_{+}^{{\rm I} \rightarrow {\rm II}}(\hbar)}{C_{-}^{{\rm I} \rightarrow {\rm II}}(\hbar)} \frac{\sqrt{2 \pi} \hbar^{-F(\hbar)}}{\Gamma(1/2 + F(\hbar) )}   \\
0  & & 1
\end{pmatrix}
\begin{pmatrix}
     \psi^{\rm II}_{+}(x,\hbar) \\
     \psi^{\rm II}_{-}(x,\hbar)
\end{pmatrix}, \\
&& \begin{pmatrix}
     \psi^{\rm II}_{+}(x,\hbar) \\
     \psi^{\rm II}_{-}(x,\hbar)
   \end{pmatrix} =
\begin{pmatrix}
 1  & & 0 \\
 i \frac{C_{-}^{{\rm II} \rightarrow {\rm III}}(\hbar)}{C_{+}^{{\rm II} \rightarrow {\rm III}}(\hbar)} \frac{\sqrt{2 \pi} e^{- \pi i F(\hbar)} \hbar^{+F(\hbar)}}{\Gamma(1/2 - F(\hbar) )} & & 1
\end{pmatrix}
\begin{pmatrix}
     \psi^{\rm III}_{+}(x,\hbar) \\
     \psi^{\rm III}_{-}(x,\hbar)
\end{pmatrix}, \\
&& \begin{pmatrix}
     \psi^{\rm III}_{+}(x,\hbar) \\
     \psi^{\rm III}_{-}(x,\hbar)
   \end{pmatrix} =
\begin{pmatrix}
  1  & & i \frac{C_{+}^{{\rm III} \rightarrow {\rm IV}}(\hbar)}{C_{-}^{{\rm III} \rightarrow {\rm IV}}(\hbar)} \frac{\sqrt{2 \pi} e^{- 2\pi i F(\hbar)} \hbar^{-F(\hbar)}}{\Gamma(1/2 + F(\hbar) )}   \\
0  & & 1
\end{pmatrix}
\begin{pmatrix}
     \psi^{\rm IV}_{+}(x,\hbar) \\
     \psi^{\rm IV}_{-}(x,\hbar)
\end{pmatrix}, 
\ee
where
\be
F(\hbar) = \sum_{n=0}^{+\infty} F_{n} \hbar^n,
\ee
and $F(\hbar)$ and $C_{\pm}^{\bullet \rightarrow \bullet}(\hbar)$ can be computed from the details  of $Q(x,\hbar)$.

$F(\hbar)$ directly has the relationship with $S_{\rm odd}(x,\hbar)$ as
\be 
F(\hbar) = \mp {\rm Res}_{x=a_\ell} S_{\rm odd}(x,\hbar), \label{eq:F_resS}
\ee
where $a_{\ell}$ is a turning point, and the sign depends on the asymptotic behavior around a turning point in the local and global coordinates.
%, whether the left or right palenel in Fig.\ref{fig:deg_web_loc}.
It can be shown as follows:
Assume that the asymptotic behavior around a turning point in the local and global Stokes graphs is the left panel in Fig.\ref{fig:deg_web_loc}.
Since
\be
\widehat{S}_{\rm odd}(y,\hbar)= \hbar^{-1} \frac{y}{2} - \frac{\kappa}{y} + O(y^{-3}),
\ee
one finds that
\be
 {\rm Res}_{y=0} \widehat{S}_{\rm odd}(y,\hbar) = - \kappa.
\ee

Suppose a neighborhood around $x=a_{\ell}$ where $y(a_{\ell},\hbar)=0$.
From eq.(\ref{eq:Sod_Sod}),
\be
{\rm Res}_{y=0} \widehat{S}_{\rm odd}(y,\hbar) = {\rm Res}_{x=a_{\ell}} S_{\rm odd}(x,\hbar),
\ee
which gives the translation of $\kappa$ in the local coordinate to $F(\hbar)$ defined in the global coordinate.
By comparing the local connection formula with the global one, Eq.(\ref{eq:F_resS}) with the minus sign can be obtained.
By repeating the same procedure, the relatinship of $F(\hbar)$ and $S_{\rm odd}(x,\hbar)$ when the asymptotic behavour around a turning point in the local and/or global coordinates change can be obtained.
As a result,  $F(\hbar) = - {\rm Res}_{x=a_\ell} S_{\rm odd}(x,\hbar)$ when the asymptotic behavior around a turning point in the local and global coordinate matches and $F(\hbar) = + {\rm Res}_{x=a_\ell} S_{\rm odd}(x,\hbar)$ if it does not.

In order to obtain $C_{\pm}(\hbar)$, we use Eqs.(\ref{eq:QQ_dW}), (\ref{eq:Sod_Sod}), and (\ref{eq:psi_tilpsi}).
From Eq.(\ref{eq:QQ_dW}) with a fixed order of $\hbar$, one finds
\be
&& \frac{y_0(x)}{2} \frac{d y_0(x)}{d x}= \pm \sqrt{Q_0(x)} = \pm S_{\rm odd,-1}(x), \nl
&& \frac{ y_0(x)^2}{2} \frac{d y_0(x)}{d x} \frac{d y_1(x)}{d x} + \left(\frac{d y_0(x)}{d x} \right)^2 \frac{y_0(x)y_1(x)}{2} - \left( \frac{dy_0(x)}{dx} \right)^2 \kappa = Q_{1}(x), \label{eq:xandy}\\
&& \qquad \vdots \nn
\ee
and recursionally solving them with $y(a_\ell,\hbar)=0$ where $a_{\ell}$ is a turning point gives the coordinate transformation.
Notice that the sign $+/-$ in the first line of Eq.(\ref{eq:xandy}) corresponds to the left/right panel in Fig.\ref{fig:deg_web_loc}.
Furthermore, by combining with (\ref{eq:Sod_Sod}) and (\ref{eq:psi_tilpsi}), one has
\be
&& C_{\pm}(\hbar) y^{\mp \kappa} \exp \left[ \pm \frac{1}{\hbar} \frac{y^2}{4} \pm \int^{y}_{\infty} dy \, \left( \widehat{S}_{\rm odd}(y,\hbar) - \frac{1}{\hbar}\widehat{S}_{\rm odd,-1}(y) + \frac{\kappa}{y} \right) \right] \nl
&=&
\begin{cases}
  \exp \left[ \pm  \frac{1}{\hbar} \int^{x}_{a_\ell} dx \, \sqrt{Q_0(x)} \pm \int^{x}_{\infty}dx \, \left( S_{\rm odd}(x,\hbar)-\frac{1}{\hbar}S_{\rm odd,-1}(x) \right)  \right]  &  \mbox{for \ $+\sqrt{Q_0(x)}$}  \\
  \exp \left[ \mp  \frac{1}{\hbar} \int^{x}_{a_\ell} dx \, \sqrt{Q_0(x)} \mp \int^{x}_{\infty}dx \, \left( S_{\rm odd}(x,\hbar)-\frac{1}{\hbar}S_{\rm odd,-1}(x) \right)  \right]  & \mbox{for \ $-\sqrt{Q_0(x)}$}
\end{cases} \label{eq:Cpm_SS}. \nl
\ee
and recursively  solve it for $C_{\pm}(\hbar)$ order by order after obtaining the coordinate transformation from eq.(\ref{eq:xandy}).
Here, we focus on $a_{\ell}=0$.
From Eq.(\ref{eq:xandy}), the 0th and 1st orders of $y$ are obtained as
\be
y_0(x) &=&
\begin{cases}
  \sqrt{2} N^{1/2} x - \frac{1}{48 \sqrt{2}} N^{5/2} x^3 + O(x^5) & \mbox{for \ $+\sqrt{Q_0(x)}$}  \\
  \sqrt{2} i N^{1/2} x - \frac{i}{48 \sqrt{2}}N^{5/2} x^3 + O(x^5) & \mbox{for \ $-\sqrt{Q_0(x)}$}  
\end{cases}, \\
y_1(x) &=&
\begin{cases}
  - \frac{1}{16 \sqrt{2}}E N^{1/2} x - \frac{3}{2048 \sqrt{2}} E N^{5/2} x^3 + O(x^5) & \mbox{for \ $+\sqrt{Q_0(x)}$}  \\
  -  \frac{i}{16 \sqrt{2}}E N^{1/2} x - \frac{3i}{2048 \sqrt{2}} E N^{5/2} x^3 + O(x^5) & \mbox{for \ $-\sqrt{Q_0(x)}$}
\end{cases},
\ee
where $\kappa$ should be chosen as $\kappa \sim F_0=\pm E/N$ to cancel divergence.
From these results the leading order of $C_{\pm}(\hbar)$ is given from Eq.(\ref{eq:Cpm_SS}) by\footnote{
In this paper, we took the asymptotic behavior around turning points as the right panel in Fig.\ref{fig:deg_web_loc}.
If one takes the left panel, the identification of $B$-cycle in Eq.(\ref{eq:Bcy_form}) gives a slightly different form.
However, by taking into account the difference of $C_\pm(\hbar)$, the final  result is unchanged.
}
\be
C_{\pm,0} &=&
\begin{cases}
  \lim_{x\rightarrow 0} \frac{\exp \left[ \pm \int^{x}_{\infty} S_{\rm odd,0}(x) \right]}{y_0(x)^{\mp \kappa}} \, = \, \exp \left[ \pm \frac{\pi i E}{2N} \right] \left( \frac{32}{N}\right)^{\pm \frac{E}{2N}} & \mbox{for \ $+\sqrt{Q_0(x)}$}  \\
  \lim_{x\rightarrow 0} \frac{\exp \left[ \mp \int^{x}_{\infty} S_{\rm odd,0}(x) \right]}{y_0(x)^{\mp \kappa}} \, = \, \left( \frac{32}{N}\right)^{\mp \frac{E}{2N}} & \mbox{for \ $-\sqrt{Q_0(x)}$}
\end{cases}.
\ee

Finally, we define the normalization and branchcut matrices as
\be
&& {\cal N}_{a_1,a_2} =
   \begin{pmatrix}
     e^{+\frac{1}{\hbar}\int^{a_2}_{a_1} dx \, S_{\rm odd,-1}(x)} && 0 \\
     0 && e^{-\frac{1}{\hbar} \int^{a_2}_{a_1} dx \, S_{\rm odd,-1}(x)}
   \end{pmatrix}, \\
   && {\cal T} =
   \begin{pmatrix}
     0 && -i \\
     -i && 0
   \end{pmatrix},
\ee
where it has only $S_{\rm odd,-1}(x)$ in the normalization matrix because the wavefunction is normalized at $x=\infty$ for $S_{{\rm odd},n>-1}(x)$, and the branchcut matrix can be defined by $S_{\rm odd}(x,\hbar) \rightarrow -S_{\rm odd}(x,\hbar)$ in the wavefunction.

\subsection{Construction of the dictionary} \label{sec:dic_A_DW}
We now  construct the dictionary translating cycles between the Airy-type and the DW-type building blocks of the Stokes graph. 

\subsubsection{$A$-cycle}
For $Q(x,\hbar) = 2 [1-\cos (N x)-E]$, the $A$-cycle around $x=0$ for the Airy-type is given by
\be
&& A=e^{\oint_{a_1}^{a_2} dx \, S_{\rm odd}(x,\hbar)}, 
\ee
where %$\pm$ depends on the first or second Riemann sheet, and
$a_{1,2}$ is the turning point given by $a_{2,1} = \pm \frac{{\rm arccos}(1-E)}{N}$ with $0<E<2$.
For simplicity, we suppose that a branch-cut connecting with both $a_{1}$ and $a_2$ as end points exists.
When replacing $E \rightarrow E \hbar$ to obtain the $A$-cycle for the DW-type, the turning points collide with each other at $x=0$.
By denoting $S^{\rm DW}_{\rm odd}(x,\hbar)$ instead of $S_{\rm odd}(x,\hbar)$ for the DW-type, which is calculated by $Q(x,\hbar) = 2 [1-\cos (N x)-E\hbar]$, the $\cal A$-cycle is defined as
\be
&& {\cal A} = e^{\oint_{|x| \ll 1} dx \, S^{\rm DW}_{\rm odd}(x,\hbar)} =: e^{\mp 2 \pi i F(\hbar)},
\ee
%Here, we took the orientation as the right panel in Fig.\ref{fig:deg_web_loc} and
where $F(\hbar)$ is given by Eq.(\ref{eq:F_resS}), and the sign depends on the asymptotic behavior around a turning point in the global coordinate.

\begin{figure}[t]
    \centering 
    \includegraphics[width=6cm]{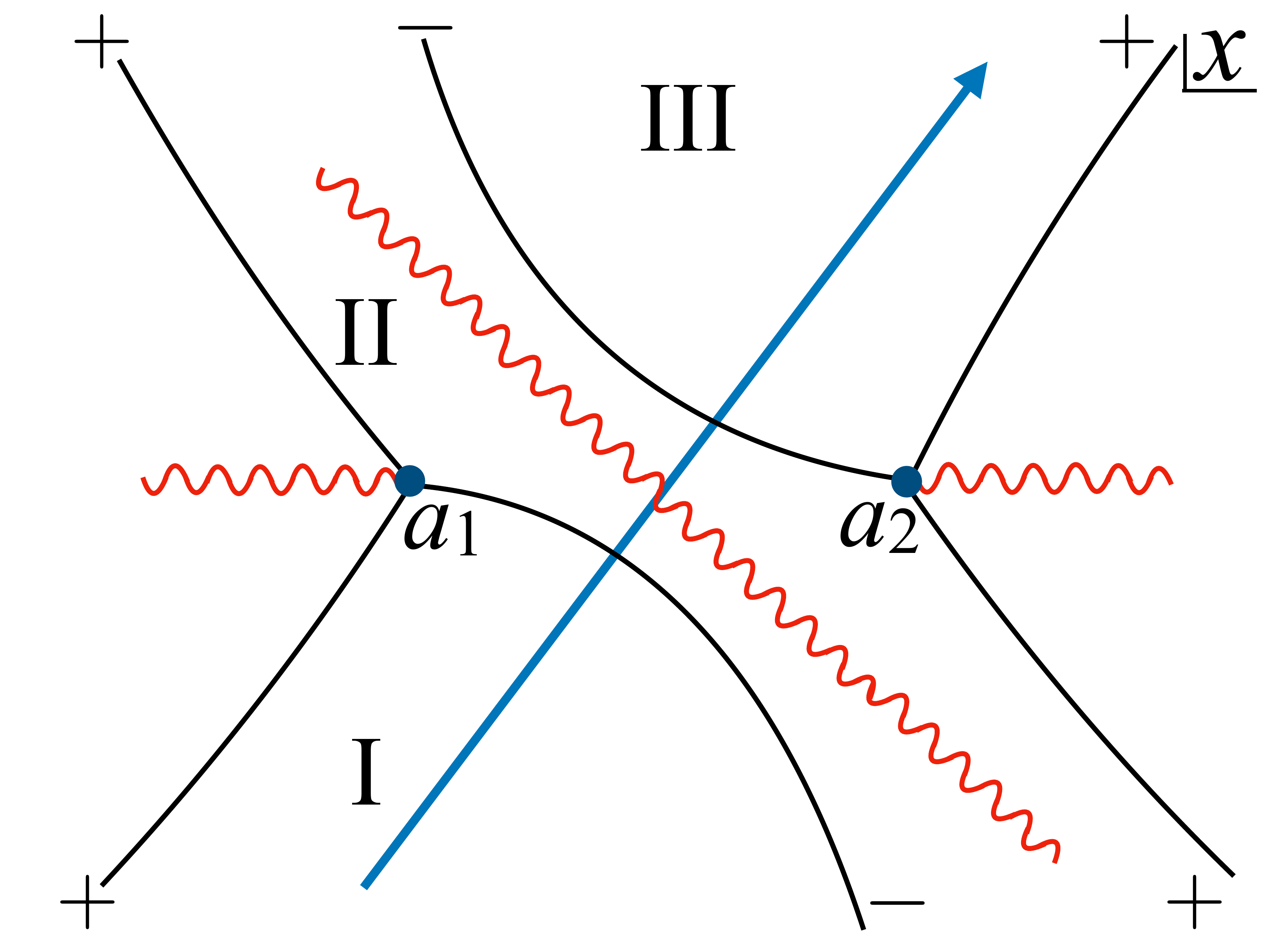}\quad\quad\quad
    \includegraphics[width=6cm]{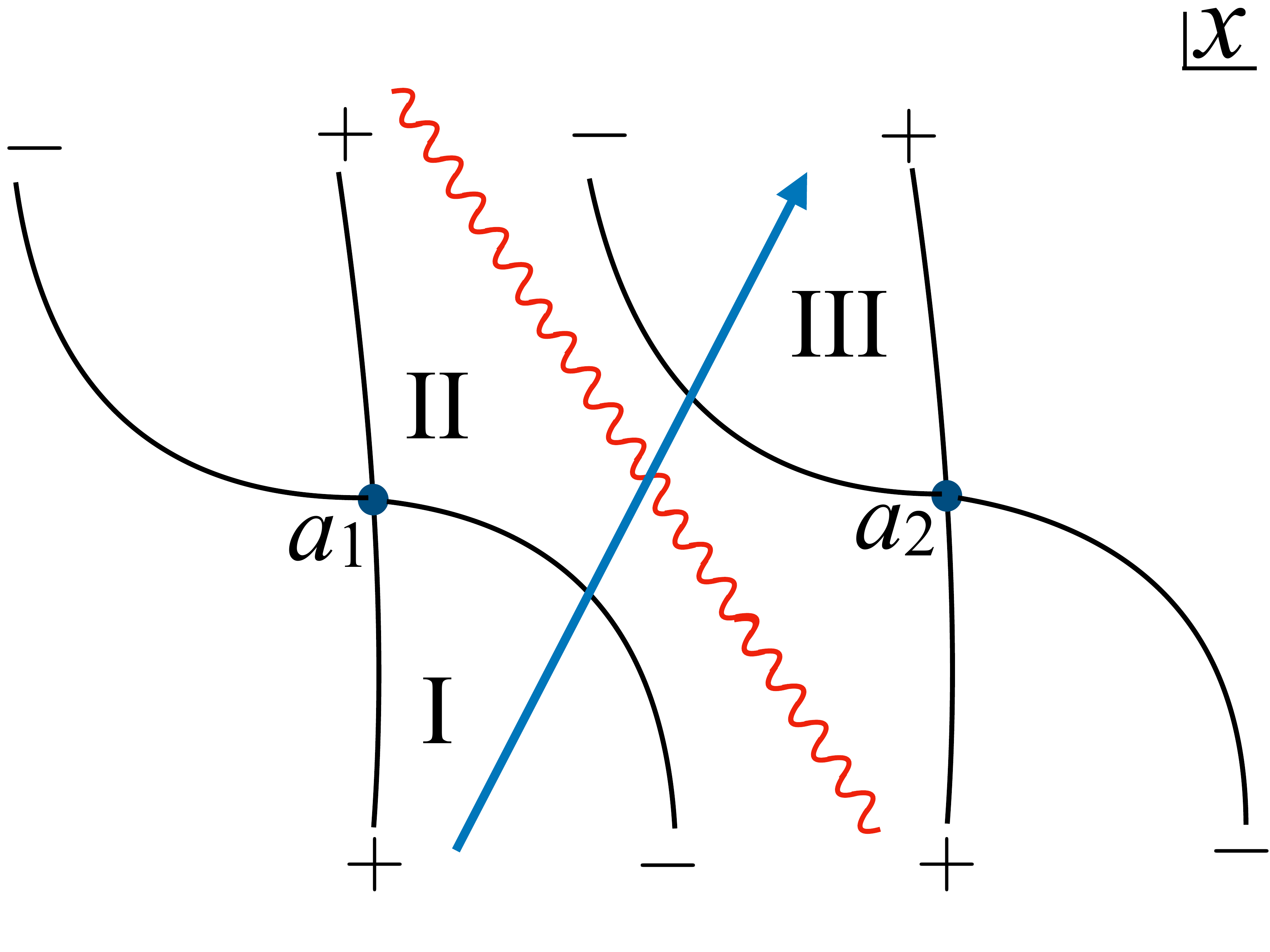}
    \caption{The orbit we consider to compare the connection formula from Airy-type (left) and the DW-type (right).}
     \label{fig:app}
\end{figure}

\subsubsection{$B$-cycle}
We consider the orbit in Fig.\ref{fig:app} and compare the monodromy matrix obtained by connection formula from the Airy-type and the DW-type.
Those are calculated by
\be
   {\rm Airy} &:& \;\; \psi_{\rm I} = M_- N_{a_1,a_2} T M^{-1}_{-} N_{a_2,a_1} \psi_{\rm III} =:D \psi_{\rm III}, \\
   {\rm DW} &:& \;\;  \psi_{\rm I} = {\cal M}^{\rm IV \rightarrow I}_{1\ominus} {\cal N}_{a_1,a_2} {\cal T}  {\cal M}^{{\rm II \rightarrow III}}_{2\ominus} {\cal N}_{a_2,a_1} \psi_{\rm III}=: {\cal D}\psi_{\rm III}.
\ee
Notice that $\psi_{\rm I}^+ = 0$ due to the asymptotic behaviour, thus $D_{21}$ and ${\cal D}_{21}$ are comparable with each other.
Those are obtained as
\be
   {\rm Airy} &:& D_{21} = -i(1+B), \\
   {\rm DW} &:& {\cal D}_{21} = -i (1+{\cal B}),
\ee
where
\be
&& B=e^{2 \int_{a_1}^{a_2} dx S_{\rm odd}(x,\hbar)}, \\
&&   {\cal B} = 2 \pi {\cal B}_0 \prod_{\ell=1}^2 \frac{C_{\ell-}(\hbar)}{C_{\ell+}(\hbar)} \frac{e^{(-1)^{\ell+1}\pi i F_\ell(\hbar)}\hbar^{F_\ell(\hbar)}}{\Gamma(1/2-F_\ell(\hbar))},  \label{eq:Bcy_form}
\ee
with ${\cal B}_0=e^{\frac{2}{\hbar} \int_{a_1}^{a_2} dx \, S^{\rm DW}_{\rm odd,-1}(x)}$, where $\ell$ is the label of turning points.

%%%%%%%%%%%%%%%%%%%%%%%%%%%%

\end{document}